\documentclass[twocolumn]{article}
\usepackage{graphicx}
\usepackage{blindtext}
\usepackage{graphicx}
\usepackage{caption}
\usepackage{abstract}
\usepackage{subcaption}
\usepackage{braket}
\usepackage{amsmath}
\usepackage{authblk}
\usepackage[percent]{overpic}

\usepackage[english]{babel}

\usepackage[letterpaper,top=2.5cm,bottom=2.5cm,left=2.5cm,right=2.5cm]{geometry}
\usepackage{amsmath}
\usepackage{amssymb}
\usepackage{graphicx}
\usepackage[colorlinks=true, allcolors=blue]{hyperref}
\usepackage{balance}

\usepackage{placeins}
\usepackage{dblfloatfix}

\title{\fontsize{14}{14}\selectfont
A Basis-Aware Approach to Quantum Sampling of the Fermi-Hubbard Ladder}

\author[1]{Adam Leicester}
\author[1]{Zixu Wang}
\author[2]{Jack Mandell}
\author[2]{Yangyang Xu}
\author[3]{Osama Muhammad Raisuddin}
\author[4]{Haimeng Zhang}
\author[1,5,*]{Jian Shi}

\affil[1]{Department of Materials Science and Engineering, Rensselaer Polytechnic Institute, Troy, New York, USA}
\affil[2]{Department of Mathematical Sciences, Rensselaer Polytechnic Institute, Troy, New York, USA}
\affil[3]{Future of Computing Institute, Rensselaer Polytechnic Institute, Troy, New York, USA}
\affil[4]{IBM Quantum, IBM T.J. Watson Research Center, Yorktown Heights, NY, USA.}
\affil[5]{Department of Physics, Rensselaer Polytechnic Institute, Troy, New York, USA}
\affil[*]{Corresponding author: shij4@rpi.edu}

\date{}

\begin{document}

\twocolumn[
\maketitle

\begin{onecolabstract}
Sample-based quantum diagonalization can reduce the classical cost of many-body calculations by restricting diagonalization to configurations identified through quantum sampling, but its efficiency depends on how compactly the target state is represented in the sampled basis. We investigate this dependence for sample-based Krylov quantum diagonalization of the half-filled Fermi-Hubbard ladder using physically motivated one-body orbital rotations. Across position, momentum, bonding/antibonding, and leg-mode representations, the most compressible basis varies systematically with interaction strength and hopping anisotropy, with momentum sampling favored in the weakly interacting regime and geometry- or interaction-adapted representations becoming advantageous as the dominant energy scale changes. Near intermediate interaction and isotropic hopping, however, even the most favorable representation remains poorly compressible. At \(U/t_{\rm leg}=5\) and \(t_{\rm rung}/t_{\rm leg}=1\), approximately \(3.5\%\) of the fixed-sector Hilbert space is required to retain \(99\%\) of the ground-state weight for a 12-site ladder, with the absolute requirement scaling approximately as \(k_{0.99}\propto 3.106^N/N\) over the computed system sizes. Statevector SKQD calculations further show that basis choice strongly affects sampling efficiency: at \(U/t=1\), momentum-basis sampling reaches a relative energy error of approximately \(0.47\%\) using only \(1.4\%\) of the fixed particle-number sector, compared with approximately \(9.5\%\) error using \(33\%\) of the sector for position-basis sampling under the same Krylov construction. Hardware calculations on systems containing up to 120 qubits retain qualitative signatures of this basis dependence while revealing increasing particle-number leakage and reconstruction sensitivity with system size. These results show that scalable sample-based diagonalization requires co-design of the sampled representation, quantum circuit, and classical reconstruction procedure.
\end{onecolabstract}

\vspace{1em}
]

\raggedbottom

\section{Introduction}

A benchmark for strongly correlated electron systems, the Fermi-Hubbard Model captures how simple local interactions give rise to complex emergent phases. It assumes a lattice of sites in which electrons interact through kinetic hopping and onsite electron-electron repulsion. Despite its compact form, the Fermi-Hubbard model serves as a testbed for rich phenomena such as charge-density order, magnetism, Mott-insulating behavior, and unconventional superconductivity \cite{arovas2022hubbard,xu2024coexistence}. As such, this model serves as an important platform for exploring condensed-matter physics and quantum simulation. However, the many-body form of this system grows combinatorially with system size, where accurate simulation is difficult for classical and pre-fault-tolerant quantum computing methods in strongly correlated regimes.

Exact classical simulations of the half-filled repulsive Fermi-Hubbard model are fundamentally limited by the exponential growth of the many-body Hilbert space with system size. Tensor-network methods, particularly the density matrix renormalization group (DMRG), provide a leading classical benchmark for calculating ground-state properties of strongly correlated one-dimensional and quasi-one-dimensional systems, including Fermi-Hubbard ladders and finite-width cylinders ~\cite{noack1996ground, leblanc2015solutions, white1992density}. DMRG represents the many-body wave function as a matrix-product state, a compressed representation that efficiently captures states with limited quantum entanglement. The accuracy of the matrix-product state is controlled by the bond dimension, $\chi$, which determines the maximum entanglement that can be represented across a division of the lattice into left and right subsystems. For a fixed target accuracy, the required bond dimension is determined primarily by the amount of entanglement, or more precisely the entanglement entropy, across this division. According to the entanglement area law, this entanglement scales primarily with the size of the boundary separating the two subsystems rather than with their respective volumes~\cite{arovas2022hubbard, stoudenmire2012studying}. For a finite-width ladder, a transverse bipartition therefore produces an entanglement boundary determined primarily by the ladder width rather than its length. Consequently, at fixed width and target accuracy, the computational cost grows approximately linearly with ladder length, whereas increasing the number of legs requires an exponentially increasing bond dimension, making wider two-dimensional systems progressively more computationally expensive to simulate. 

Quantum Monte Carlo (QMC) methods provide yet another powerful classical approach to the half-filled Fermi-Hubbard model on bipartite lattices (e.g., square or honeycomb), where particle-hole symmetry eliminates the fermion sign problem. In this setting, a bipartite lattice can be divided into two sublattices such that nearest-neighbor hopping occurs only between opposite sublattices, as in square, ladder, or cubic geometries. Under these conditions, QMC has enabled simulations of three-dimensional Hubbard systems containing up to 8000 sites~\cite{sun2025boosting}. However, this efficiency relies on the absence of a sign problem: when Monte Carlo configurations acquire negative or complex weights, cancellations between samples cause the statistical error to grow rapidly, often exponentially, with system size or inverse temperature. For this reason, QMC becomes much less applicable in generalized Hubbard models with elements like next-nearest-neighbor hopping, geometric frustration, or doping away from half-filling~\cite{arovas2022hubbard}.

Quantum simulation offers an alternative route for studying Fermi-Hubbard physics, but near-term implementations face bottlenecks that differ from those of classical methods. In gate-based simulations, fermionic degrees of freedom must be mapped onto qubits, and this mapping can turn local fermionic terms into nonlocal qubit operations. As a result, current implementations are often limited by circuit depth, qubit connectivity, noise, and the need for error-mitigation techniques~\cite{jafarizadeh2025recipe}. 

Recent quantum computing progress illustrates several promising directions. Variational quantum algorithms have been used to estimate ground-state properties of \(8\times 1\) and \(4\times 2\) Fermi-Hubbard systems using 16 qubits, demonstrating qualitative signatures such as metal-insulator behavior, Friedel oscillations, and antiferromagnetic correlations~\cite{stanisic2022observing}. Larger-scale digital simulations have focused primarily on real-time dynamics: IBM superconducting processors have been used to simulate one-dimensional Fermi-Hubbard dynamics with more than 100 qubits, while Google's Willow processor has been used to simulate two-dimensional Fermi-Hubbard dynamics on lattices up to \(6\times6\), corresponding to 72 qubits~\cite{chowdhury2026quantum, alam2025programmable}. These demonstrations show rapid progress in hardware scale and circuit construction for Hubbard-model simulation, but they also highlight the remaining challenges of obtaining accurate ground-state energies for strongly correlated two-dimensional systems on noisy, pre-fault-tolerant devices.

Sample-based quantum diagonalization (SQD) is a hybrid quantum-classical approach in which a quantum processor is used to identify occupation-number configurations carrying significant ground-state weight~\cite{robledo2025chemistry, kanno2026quantum, raisuddin2025promise}. These sampled configurations define a reduced subspace in which the Hamiltonian can be diagonalized classically. SQD can additionally use known symmetries of the target problem to correct or replace sampled bitstrings that fall outside the desired symmetry sector through a post-processing procedure known as configuration recovery~\cite{robledo2025chemistry}. Extensions of SQD have already been applied to the Fermi-Hubbard model, including ground-state estimation for \(8\times2\) and \(28\times1\) systems~\cite{nogaki2025symmetry,yeter2026towards}. Sample-based Krylov quantum diagonalization (SKQD) extends this framework by sampling a family of Trotterized time-evolved states generated from an initial reference state, thereby exploring configurations beyond those populated by a single reference or variational state~\cite{yu2025quantum, misciasci2026evaluating, kirby2026observation}. SKQD has been demonstrated on the single-impurity Anderson model, a related model of strongly correlated electron physics~\cite{yu2025quantum}.

A central bottleneck for both SQD and SKQD is the assumption that the target ground state is sufficiently compressible in the sampled basis~\cite{robledo2025chemistry, yu2025quantum, gaberle2026critical}. In this case, an accurate reduced subspace can be constructed from a relatively small number of basis states without losing significant ground-state weight. When the ground state is distributed over many basis states, the sampled subspace must grow substantially, reducing the advantage of the diagonalization procedure. Recent analyses of SQD applied to Heisenberg and Hubbard models have shown that the number of computational-basis configurations required to achieve fixed energy accuracy can grow exponentially with system size even under optimal configuration selection, indicating that this limitation can originate from the intrinsic distribution of the many-body wave function rather than from sampling inefficiency alone~\cite{gaberle2026critical}. At the same time, the determinant support relevant to sample-based diagonalization is basis dependent, motivating the use of orbital rotations to search for representations in which the physically relevant wave function is concentrated over fewer configurations~\cite{bonillavargas2026dynamical}.

SKQD introduces an additional challenge because its circuit depth depends on both the preparation of the reference state and the complexity of the Trotterized Krylov evolution. As a result, the potential sampling advantages of SKQD may be offset by the additional noise accumulated in deeper circuits. These issues impose natural limitations for the Fermi-Hubbard model, particularly in intermediate-interaction regimes where sparse representations of the ground state are often difficult to identify but where much of the model's rich correlated physics emerges. On real quantum hardware, limited qubit connectivity and the overhead of implementing fermionic hopping terms further increase the circuit depth required for Krylov evolution, causing errors to accumulate more severely as the system size grows. 

In this work, we investigate structure- and parameter-aware basis rotations for SKQD applied to the Fermi-Hubbard model. The goal is to identify orbital bases in which the ground-state wave function is more compressible, allowing the sampled subspace to capture relevant low-energy physics with fewer basis states. Within a quantum-circuit implementation, these rotations can improve reference-state preparation, alter the measurement basis, and affect the basis state distribution sampled by Trotterized Krylov states. We demonstrate this approach on the Fermi-Hubbard ladder, a quasi-one-dimensional system that is tractable with DMRG. By comparing site, momentum basis, referred to below as the momentum basis for brevity, bonding/antibonding, and leg-mode representations, we evaluate the extent to which physically motivated one-body orbital rotations can improve SKQD sampling efficiency and where such rotations fail to overcome the intrinsic density of the correlated ground state.

\section{Methods}

\begin{figure}[!t]
    \centering
    \includegraphics[width=\columnwidth]{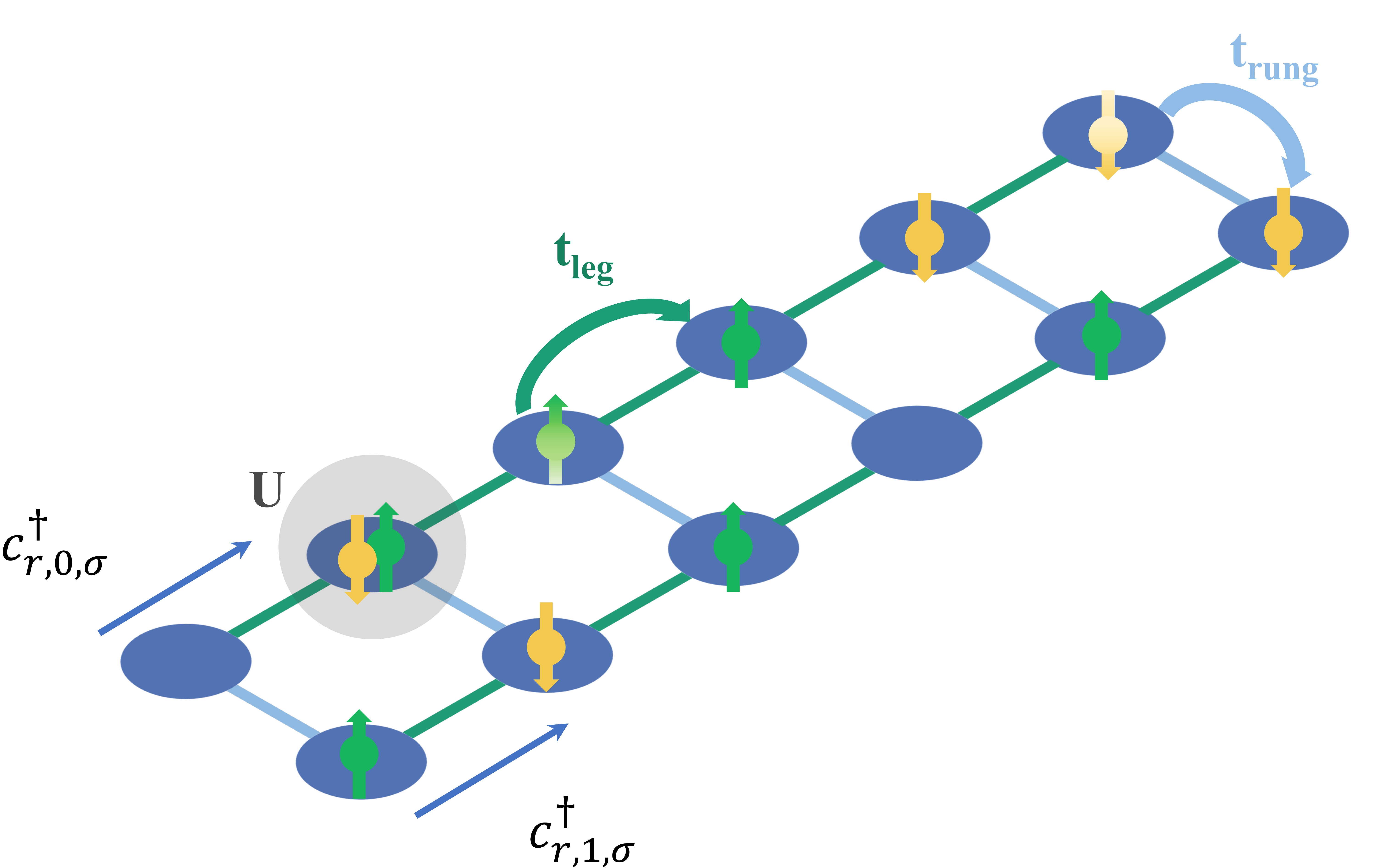}
    \caption{Two-leg Fermi-Hubbard ladder considered in this work. Fermions hop along each leg with amplitude $t_{\mathrm{leg}}$ and between legs along each rung with amplitude $t_{\mathrm{rung}}$. Opposite-spin particles occupying the same site experience the onsite interaction $U$. The creation operators $\hat{c}^\dagger_{r,0,\sigma}$ and $\hat{c}^\dagger_{r,1,\sigma}$ label the two sites on rung $r$.}
    \label{fig:LadderModel}
\end{figure}

The two-leg Fermi-Hubbard ladder considered here contains \(N_{\rm rung}\) total rungs, or equivalently the sites on each leg, giving \(N=2N_{\rm rung}\) spatial sites in total (Fig.~\ref{fig:LadderModel}). This geometry provides a minimal model for correlated ladder physics relevant to ladder cuprate materials such as \(Sr_{n-1}Cu_{n+1}O_{2n}\), while remaining tractable with DMRG calculations~\cite{noack1996ground,dagotto1999experiments,jeckelmann1998comparison}. The general Fermi-Hubbard Hamiltonian is

\begin{equation} \label{eq:SiteHamiltonian}
\hat{H}
=
- \sum_{\langle i,j\rangle,\sigma}
t_{ij}\left(
\hat{c}^{\dagger}_{i\sigma}\hat{c}_{j\sigma}
+
\hat{c}^{\dagger}_{j\sigma}\hat{c}_{i\sigma}
\right)
+
U \sum_i
\hat{n}_{i\uparrow}\hat{n}_{i\downarrow}.
\end{equation}
where \(U\) denotes the onsite Coulomb repulsion between two electrons of opposite spin occupying the same lattice site, and \(t_{ij}\) is the hopping amplitude between sites, \(i\) and \(j\). The operator \(\hat{c}^{\dagger}_{i\sigma}\) creates a fermion with spin \(\sigma\) on site \(i\), while \(\hat{c}_{i\sigma}\) annihilates one.  The corresponding number operator is \(\hat{n}_{i\sigma}=\hat{c}^{\dagger}_{i\sigma}\hat{c}_{i\sigma}\), so that \(\hat{n}_{i\uparrow}\hat{n}_{i\downarrow}\) counts doubly occupied sites. 

We apply this Hamiltonian to a two-leg Fermi-Hubbard ladder, such as the \(6\times2\) ladder modeled in Fig.~\ref{fig:LadderModel}. We label each site by a rung index \(r=1,\ldots,N_{\rm rung}\), a leg index \(\ell\in\{0,1\}\), and a spin index \(\sigma\). The specialized ladder Hamiltonian is

\begin{equation}
\label{eq:LadderHamiltonian}
\begin{aligned}
\hat{H}_{\rm ladder}
=&
-t_{\rm leg}
\sum_{r=1}^{N_{\rm rung}-1}
\sum_{\ell=0}^{1}
\sum_{\sigma}
\left(
\hat{c}^{\dagger}_{r,\ell,\sigma}
\hat{c}_{r+1,\ell,\sigma}
+
\mathrm{h.c.}
\right)
\\
&-
t_{\rm rung}
\sum_{r=1}^{N_{\rm rung}}
\sum_{\sigma}
\left(
\hat{c}^{\dagger}_{r,0,\sigma}
\hat{c}_{r,1,\sigma}
+
\mathrm{h.c.}
\right)
\\
&+
U
\sum_{r=1}^{N_{\rm rung}}
\sum_{\ell=0}^{1}
\hat{n}_{r,\ell,\uparrow}
\hat{n}_{r,\ell,\downarrow}.
\end{aligned}
\end{equation}
Here, \(t_{\rm leg}\) describes nearest-neighbor hopping along each leg of the ladder, while \(t_{\rm rung}\) describes hopping across each rung. We investigate open boundary conditions along the ladder directions, so the leg-hopping sum runs only from \(r=1\) to \(N_{\rm rung}-1\). Rung hopping is treated similarly without additional periodic wrapping.

Although the site basis is the natural representation of the onsite Hubbard interaction, it is not necessarily the most compact representation of the ground-state wave function. We therefore consider one-body orbital rotations~\cite{helgaker2000orbital, kivlichan2018quantum} that transform the fermionic creation operators according to

\begin{equation}
\label{eq:oneBodyRotation}
\hat{a}^{\dagger}_{p\sigma}
=
\sum_i R_{ip}\hat{c}^{\dagger}_{i\sigma},
\end{equation}
where \(\hat{c}^{\dagger}_{i\sigma}\) creates a fermion with spin \(\sigma\) on site \(i\) in the original site basis, while \(\hat{a}^{\dagger}_{p\sigma}\) creates a fermion with spin \(\sigma\) in the rotated orbital \(p\) within the rotated basis. The unitary orbital-rotation matrix, \(R\), whose columns contain the coefficients \(R_{ip}\), is an \(N\times N\) matrix that relates the original basis and the rotated basis. Here \(N\) is the number of spatial orbitals, and the same orbital rotation is applied independently within each spin sector. Under this transformation, the one-body hopping matrix transforms as \(h' = R^{\dagger}hR\), while the onsite interaction generally becomes nonlocal in the rotated basis. In this work, we compare four orbital representations. The first is the position (site) basis, \(\mathcal{B}_{\rm site}=\{|r,\ell,\sigma\rangle\}\), in which \(r\) labels the rung and \(\ell\in\{0,1\}\) labels the leg. The second is the momentum basis, \(\mathcal{B}_{\rm mom}=\{|p,\sigma\rangle\}\), obtained by diagonalizing the complete one-body hopping matrix \(h\). The third is the structure-aware bonding/antibonding basis, \(\mathcal{B}_{\rm BA}=\{|r,+,\sigma\rangle,|r,-,\sigma\rangle\}\), where \(+\) and \(-\) denote bonding and antibonding modes defined locally on each rung. The fourth is the leg-mode basis, \(\mathcal{B}_{\rm leg}=\{|\ell,p,\sigma\rangle\}\), obtained by diagonalizing the intra-leg hopping independently along each leg while retaining the leg index \(\ell\)~\cite{lin1997nchain}.

For rung \(r\), the bonding and antibonding creation operators are defined as

\begin{align}
\hat{b}^{\dagger}_{r,+,\sigma}
&=
\frac{1}{\sqrt{2}}
\left(
\hat{c}^{\dagger}_{r,0,\sigma}
+
\hat{c}^{\dagger}_{r,1,\sigma}
\right),
\\[0.5em]
\hat{b}^{\dagger}_{r,-,\sigma}
&=
\frac{1}{\sqrt{2}}
\left(
\hat{c}^{\dagger}_{r,0,\sigma}
-
\hat{c}^{\dagger}_{r,1,\sigma}
\right).
\end{align}
Equivalently, this local rung rotation can be written as 

\begin{equation}
\label{eq:bAbRotation}
\begin{pmatrix}
\hat{b}^{\dagger}_{r,+,\sigma} \\
\hat{b}^{\dagger}_{r,-,\sigma}
\end{pmatrix}
=
R_{\rm BA}
\begin{pmatrix}
\hat{c}^{\dagger}_{r,0,\sigma} \\
\hat{c}^{\dagger}_{r,1,\sigma}
\end{pmatrix},
\qquad
R_{\rm BA}
=
\frac{1}{\sqrt{2}}
\begin{pmatrix}
1 & 1 \\
1 & -1
\end{pmatrix}.
\end{equation}
A two-site derivation of this transformation, together with its extension into the ladder model and its relation to more general correlated many-body transformations, is provided in Sec.~\ref{sec:TwoSiteBasisAnalysis} of the Supplementary Information.

Natural-orbital representations, obtained by diagonalizing the one-body reduced density matrix, were also briefly considered as a state-adapted one-body basis as they have previously been employed in SKQD calculations of the single-impurity Anderson model \cite{yu2025quantum}. For the representative isotropic intermediate-interaction regime examined here, the natural-orbital basis did not provide a systematic advantage over the physically motivated representations considered above and generally produced results comparable to those obtained in the momentum basis. Because its construction additionally requires prior information about the interacting state, we therefore focus on orbital bases that can be defined directly from the Hamiltonian and lattice geometry.

The momentum basis is obtained by diagonalizing the one-body hopping terms in Eq.~\eqref{eq:SiteHamiltonian}, so that the single-particle modes, labeled by \(p\), are organized by their kinetic energy rather than by their real-space position on the ladder, \((r,\ell)\). In contrast, the bonding/antibonding basis diagonalizes only the local rung degree of freedom, while retaining a real-space description along the ladder direction. These bases, \[\mathcal{B}\in\left\{\mathcal{B}_{\rm site},\mathcal{B}_{\rm mom},\mathcal{B}_{\rm BA},\mathcal{B}_{\rm leg}\right\},\] therefore expose complementary physical structures of the Fermi-Hubbard ladder and allow us to test whether basis choice improves ground-state compressibility and the quality of SKQD sampling.

To compare basis-dependent compressibility, exact diagonalization (ED) was performed for systems with up to \(7\times2\) sites. Reference ground-state energies for systems beyond those accessible to exact diagonalization were obtained using DMRG; bond-dimension convergence and truncation-error diagnostics are provided in Sec.~\ref{sec:DMRGReference} of the Supplementary Information. The compressibility of the ground state is closely related to the structure of the Hamiltonian in the chosen occupation-number basis. When a basis diagonalizes or simplifies the dominant terms of the Hamiltonian, fewer configurations may be strongly coupled and the ground-state weight can become concentrated over a smaller subset of determinants. In contrast, broadly distributed off-diagonal couplings can generate weight across a larger configuration space. Guided by this intuition, we use the ED ground state to compute the weight of each occupation-number basis state in a chosen orbital representation and determine the minimum number of basis states required to capture a specified fraction of the total ground-state weight. 

For a chosen orbital representation, $\mathcal{B}$, an occupation-number basis state is written as

\begin{equation}
\label{eq:OccupationBasisState}
|\mathbf{n}\rangle_{\mathcal{B}}
=
\prod_{p}
\left(
\hat{a}^{\dagger}_{p\uparrow}
\right)^{n_{p\uparrow}}
\prod_{p}
\left(
\hat{a}^{\dagger}_{p\downarrow}
\right)^{n_{p\downarrow}}
|0\rangle,
\end{equation}
where $\hat{a}^{\dagger}_{p\sigma}$ creates a fermion with spin $\sigma$ in orbital $p$, $n_{p\sigma}\in\{0,1\}$ specifies whether the corresponding spin orbital is unoccupied or occupied, and the products are taken in a fixed canonical orbital ordering. For the half-filled $S_z=0$ systems considered here, the occupation numbers satisfy

\begin{equation}
\label{eq:HalfFillingConstraint}
\sum_{p} n_{p\uparrow}
=
\sum_{p} n_{p\downarrow}
=
\frac{N}{2},
\end{equation}
where $N$ is the number of spatial orbitals. For example, in a six-orbital-momentum representation, the configuration

\begin{equation}
|111000\rangle_{\mathrm{mom},\uparrow}
\otimes
|111000\rangle_{\mathrm{mom},\downarrow}
\end{equation}
corresponds to occupation of the first three orbitals in each spin sector ordered by increasing single-particle kinetic energy.

Let \(|\Psi_0\rangle\) denote the exact ground state of \(\hat H\). In representation \(\mathcal B\), it can then be expanded as

\begin{equation}
\label{eq:GroundStateExpansion}
|\Psi_0\rangle
=
\sum_{\mathbf{n}}
\alpha_{\mathbf{n}}^{(\mathcal{B})}
|\mathbf{n}\rangle_{\mathcal{B}},
\end{equation}
where

\begin{equation}
\label{eq:GroundStateOverlap}
\alpha_{\mathbf{n}}^{(\mathcal{B})}
=
{}_{\mathcal B}\langle\mathbf{n}|\Psi_0\rangle
\end{equation}
is the overlap amplitude between the selected occupation-number configuration and the ground state. The corresponding ground-state weight is

\begin{equation}
\label{eq:BasisStateWeight}
w_{\mathbf{n}}^{(\mathcal{B})}
=
\left|
\alpha_{\mathbf{n}}^{(\mathcal{B})}
\right|^2,
\end{equation}
with

\begin{equation}
\sum_{\mathbf{n}}
w_{\mathbf{n}}^{(\mathcal{B})}
=
1.
\end{equation}
After sorting the configurations in descending order of weight,

\begin{equation}
w_{(1)}^{(\mathcal B)}
\geq
w_{(2)}^{(\mathcal B)}
\geq
\cdots,
\end{equation}
we define the compressibility measure

\begin{equation}
\label{eq:CompressibilityDefinition}
k_W^{(\mathcal B)}
=
\min
\left\{
k:
\sum_{j=1}^{k}
w_{(j)}^{(\mathcal B)}
\geq W
\right\},
\end{equation}
where $k_W^{(\mathcal B)}$ is the minimum number of occupation-number configurations required to retain a cumulative ground-state weight $W$. Unless otherwise specified, we use $W=0.99$.

\begin{figure}[!t]
    \centering
    \includegraphics[width=\columnwidth]{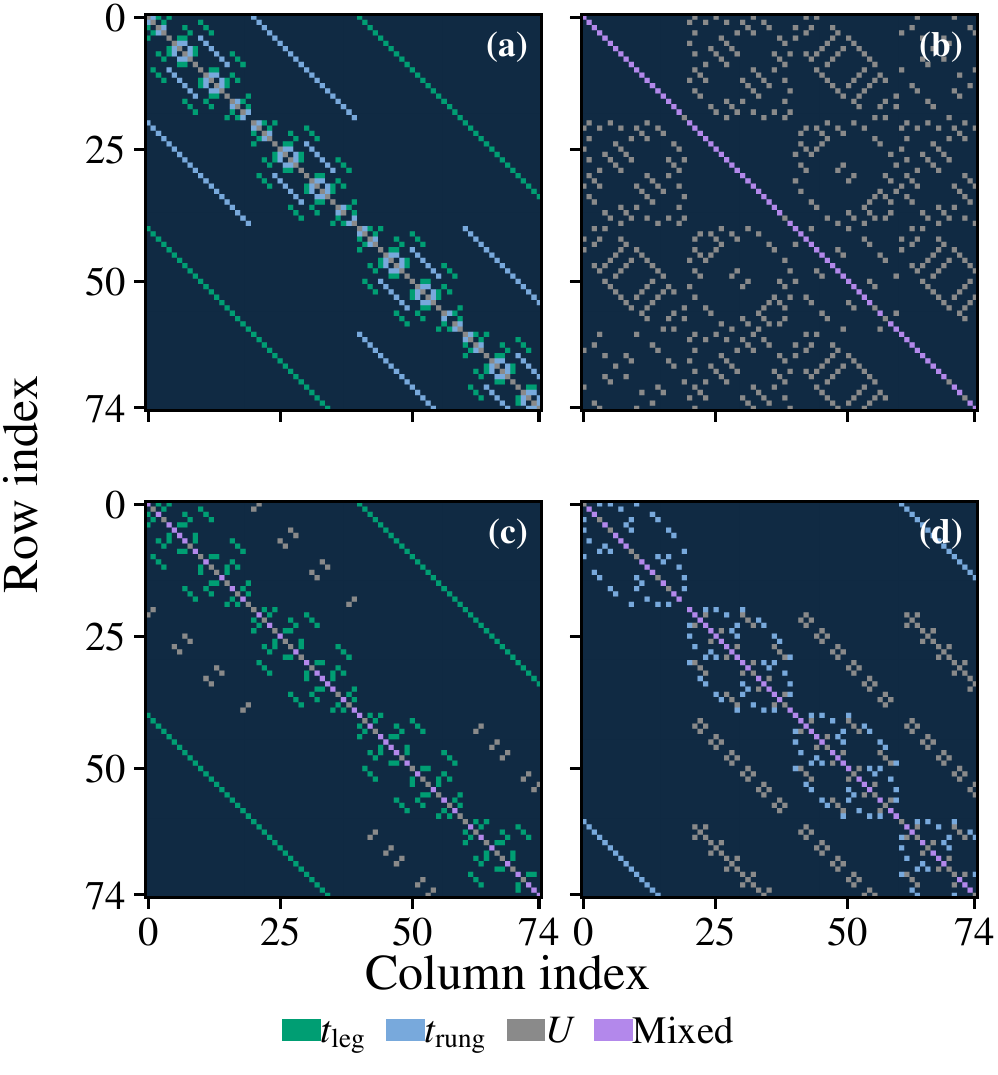}
    \caption{Matrix structure of the six-site Fermi-Hubbard Hamiltonian, showing the first $75\times75$ basis-state indices in the (a) site, (b) momentum, (c) bonding/antibonding, and (d) leg-mode representations. Matrix elements are colored according to contributions from $t_{\mathrm{leg}}$, $t_{\mathrm{rung}}$, $U$, or mixed terms.}
    \label{fig:FHMatrixHamiltonian}
\end{figure}

The same occupation-number configurations define the matrix representation of the Hamiltonian in orbital basis $\mathcal{B}$ through

\begin{equation}
\label{eq:HamiltonianBasisMatrix}
H_{\mathbf{n}'\mathbf{n}}^{(\mathcal B)}
=
{}_{\mathcal B}\langle\mathbf{n}'|
\hat{H}
|\mathbf{n}\rangle_{\mathcal B}.
\end{equation}
Examining this matrix provides complementary intuition for the observed basis-dependent compressibility by revealing how the diagonal and off-diagonal couplings between occupation-number configurations change under the orbital transformation. The upper-left \(75\times 75\) entries of the representative Hamiltonian matrices for the half-filled \(3\times2\) ladder are shown in Fig.~\ref{fig:FHMatrixHamiltonian}, illustrating how different orbital bases can expose different physical structures of the model. The full Hamiltonian in the fixed half-filled \(S_z=0\) sector has dimension \(400\times400\) in which the corresponding Hilbert-space dimension is derived in Sec.~\ref{sec:HilbertSpaceDimension} of the Supplementary Information. The six-site \(3\times2\) ladder is the smallest ladder geometry in which the bonding/antibonding and leg-mode transformations are structurally inequivalent: the former diagonalizes hopping across two-site rungs, whereas the latter diagonalizes hopping along three-site legs.

In these matrices, the horizontal and vertical axes denote column and row indices of the Hamiltonian matrix in the corresponding many-body basis. Green and blue indicate matrix elements containing only hopping contributions proportional to \(t_{\mathrm{leg}}\) and \(t_{\mathrm{rung}}\) in Eq.~\eqref{eq:LadderHamiltonian} respectively, while gray entries indicate matrix elements containing only onsite interaction contributions proportional to \(U\). Purple entries indicate matrix elements that contain both hopping and onsite-interaction contributions. After basis transformation, the nonzero matrix elements are not generally restricted to integer multiples of the bare hopping and interaction parameters.

To build intuition for the basis comparison, it is useful to consider two limiting cases of the Hubbard ladder. In the site basis, the onsite interaction term in Eq.~\eqref{eq:SiteHamiltonian} \[U\sum_i \hat{n}_{i\uparrow}\hat{n}_{i\downarrow}\] is diagonal in the occupation-number basis. In the limit \(t_{\mathrm{rung}}=t_{\mathrm{leg}}=0\) and \(U>0\), all states with no double occupation have zero interaction energy, while configurations with double occupation are penalized by an energy cost \(U\) per doubly occupied site. Thus, the site basis directly exposes the no-doublon structure of the large-\(U/t\) regime, where \(t\) describes the isotropic  \(t_{\mathrm{rung}}=t_{\mathrm{leg}}=t=1\) case. Figure~\ref{fig:FHMatrixHamiltonian}(a) shows the Hamiltonian matrix in the site basis. In the strict zero-hopping limit, the hopping contribution vanishes and the Hamiltonian becomes fully diagonal in this basis. This makes the site basis the natural representation for interaction-dominated physics.

For finite hopping at large \(U/t\), however, the ground state is generally not a single occupation-number basis state. Instead, electrons can virtually hop onto neighboring occupied sites and return, generating effective spin exchange within the no-doublon subspace, which contains configurations such as \[\ket{101010}_{\mathrm{site},\uparrow}\otimes \ket{010101}_{\mathrm{site},\downarrow}.\] Here, the basis states are organized by rung-major indexing, such that the organization follows \(\ket{(1,0)(1,1)(2,0)(2,1)(3,0)(3,1)}_{\mathrm{site},\sigma}\) with each coordinate following \((r,\ell)\). In the fixed \(S_z=0\) sector, this no-doublon subspace contains \[\binom{N}{N/2}=\frac{N!}{(N/2)!(N/2)!}\] spin configurations, where \(N\) is the number of lattice sites. Therefore, even though the site basis naturally captures the suppression of double occupation, the spin subspace relevant to sampling-based procedures such as SKQD still grows combinatorially with system size.

Conversely, in the noninteracting limit \(U=0\), the Hamiltonian is entirely one-body and can be diagonalized by transforming to the momentum basis. Figure~\ref{fig:FHMatrixHamiltonian}(b) shows the Hamiltonian matrix in this basis. In the strict \(U=0\) limit, the onsite interaction vanishes and the many-body Hamiltonian becomes diagonal in the occupation-number basis of the hopping eigenmodes. This makes the momentum basis the natural representation for hopping-dominated or weakly correlated physics. 

In this basis the hopping term takes the form \[\hat{H}_t = \sum_{p,\sigma} \epsilon_p\hat{a}^{\dagger}_{p\sigma}\hat{a}_{p\sigma},\] where \(\epsilon_p\) is the single-particle kinetic energy of mode \(p\). The many-body ground state is then a filled Fermi sea, obtained by occupying the lowest-energy single-particle modes for each spin sector, up to possible degeneracies at the Fermi level. For a half-filled six-site model with \(N_\uparrow=N_\downarrow=3\), and with kinetic modes ordered from lowest to highest energy, this corresponds to the occupation-number basis state \[\ket{111000}_{\mathrm{mom},\uparrow}\otimes\ket{111000}_{\mathrm{mom},\downarrow}.\] Thus, the momentum basis provides a compact representation in the weakly interacting regime.

The bonding/antibonding basis provides an intermediate, structure-aware rotation.  As shown in Fig.~\ref{fig:FHMatrixHamiltonian}(c), this basis produces a mixed Hamiltonian structure between the site and momentum representations: neither the hopping nor onsite interaction terms are fully diagonal in the many-body occupation basis. Its utility, however, is demonstrated by diagonalizing the local rung degree of freedom while retaining a real-space description along the ladder direction. The bonding/antibonding basis is therefore physically natural in the strong-rung regime, \(t_{\rm rung}\gg t_{\rm leg}\). Conversely, the leg-mode basis demonstrated in Fig. ~\ref{fig:FHMatrixHamiltonian}(d) diagonalizes the intra-leg hopping structure while retaining the real-space distinction between the two legs, making this description effective in describing the weak-rung \(t_{\mathrm{rung}}\ll t_{\mathrm{leg}}\) regime. Together, these limiting cases motivate the use of basis-dependent compressibility as a diagnostic for identifying which orbital basis provides the most compact representation of the interacting ground state.

These limiting cases show that basis choice can strongly affect the compactness of the ground-state representation. The most challenging regime is the isotropic intermediate-interaction case, \(t_{\rm rung}=t_{\rm leg}\) and \(U=O(t)\), where none of these simple limiting representations is expected to be clearly optimal. Although one could in principle optimize orbital rotations variationally using SKQD performance or ground-state compressibility as an objective function, such optimization is beyond the scope of this work. Instead, we compare the physically motivated site, kinetic, bonding/antibonding, and leg-mode bases.

For the remainder of this work, we investigate the quality of SKQD sampling on the Fermi-Hubbard ladder model using both statevector and quantum-hardware calculations on systems containing up to \(30\times2\) sites. As defined above, \(k_W^{(\mathcal B)}\) denotes the minimum number of occupation-number configurations required to capture a cumulative ground-state weight \(W\), with \(W=0.99\) used throughout unless otherwise specified. To compare compressibility across systems with different Hilbert-space dimensions, we additionally define the normalized compressibility fraction 

\begin{equation}
f_W^{(\mathcal B)}=
\frac{k_W^{(\mathcal B)}}
{\dim\mathcal{H}_{N_\uparrow,N_\downarrow}},
\end{equation} 
where \(\dim\mathcal{H}_{N_\uparrow,N_\downarrow}\) is the dimension of the fixed particle-number and spin sector. A more compressible representation therefore corresponds to smaller values of both \(k_W^{(\mathcal B)}\) and \(f_W^{(\mathcal B)}\), with the latter allowing direct comparison across different system sizes.

For the SKQD calculations, Krylov states were generated through repeated Trotterized time evolution of an initial reference state, with the resulting measurement distributions used to identify occupation-number configurations for reduced-subspace diagonalization. Unless otherwise specified, the statevector and hardware calculations use Krylov states with up to Krylov dimension \(K+1=8\), where \(K\) is the maximum number of Trotter steps with time increments of \(\Delta t=0.025\). A detailed description of the reference-state preparation, orbital rotations, and Trotterized Krylov evolution is provided in Sec.~\ref{sec:SKQDCircuitPreparation}, while the FSWAP-based hardware implementation is described in Sec.~\ref{sec:FSWAPNetwork} of the Supplementary Information.

\section{Results}

\subsection{Basis-Dependent Ground-State Compressibility}

\begin{figure*}[!t]
    \centering
    \includegraphics[width=.92\textwidth]{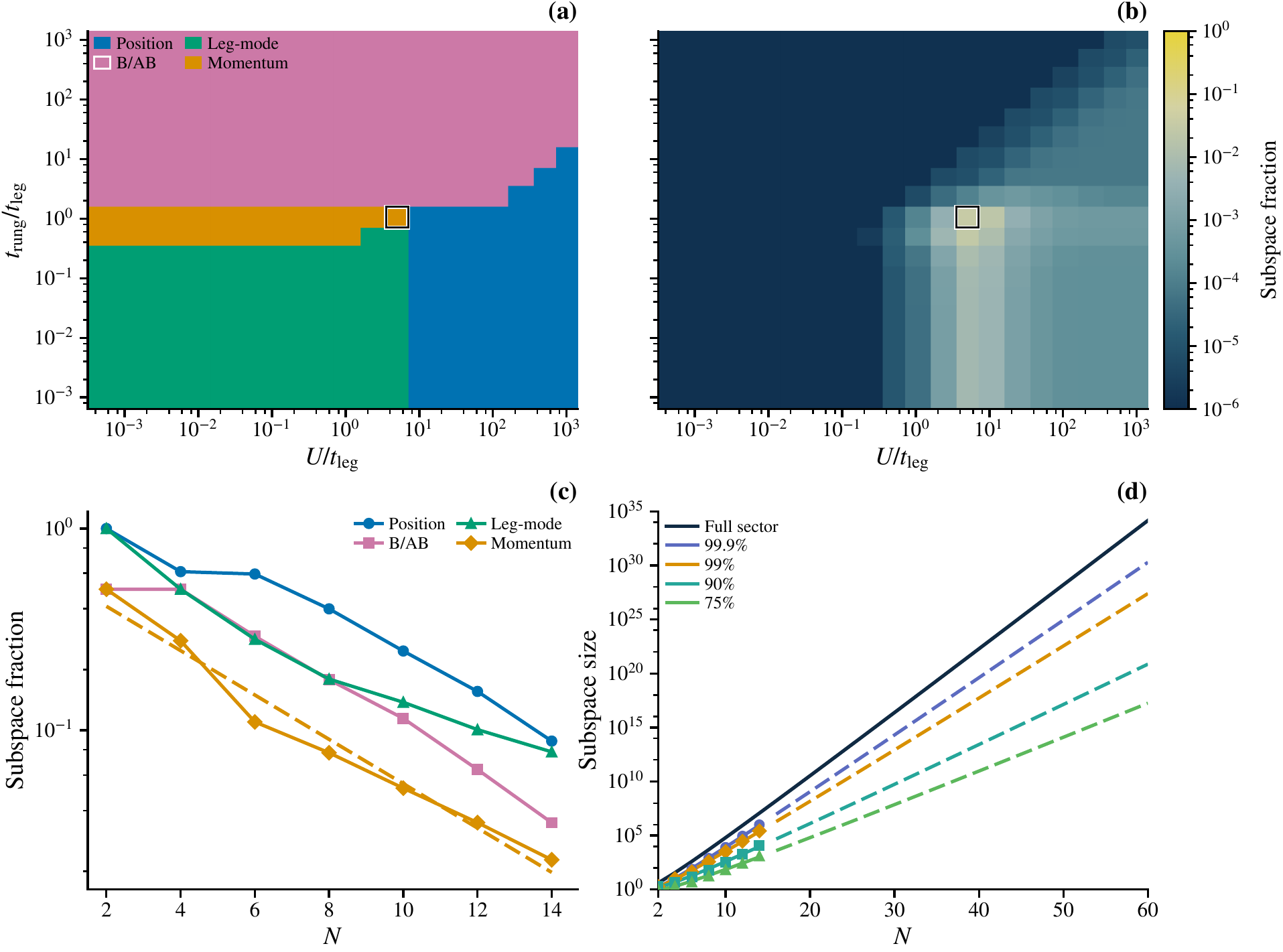}
    \caption{Basis-dependent compressibility \(f_W\) and size scaling of the half-filled Fermi-Hubbard ladder. (a) Most compressible representation across the sampled parameter space for a 12-site ladder at $W\geq0.99$. (b) Minimum required subspace fraction, with the open square marking the representative least-compressible point. (c) Subspace fraction \(f_W\) versus $N$ at this point for the four considered bases; the dashed curve is the Stirling-motivated $99\%$ fit. (d) Subspace-size scaling for several cumulative-weight thresholds from the momentum basis, with dashed curves denoting extrapolation beyond the computed system sizes.}
    \label{fig:12SiteCompressibility}
\end{figure*}

To quantify the intrinsic sparsity available to sample-based diagonalization, we use exact diagonalization to determine the minimum number of occupation-number basis states required to capture \(99\%\) of the ground-state weight. Figure~\ref{fig:12SiteCompressibility}(a) shows the most compressible representation across the sampled \((U/t_{\rm leg},t_{\rm rung}/t_{\rm leg})\) parameter space of the half-filled 12-site ladder. The preferred representation varies strongly with the underlying physical regime. In the weakly interacting limit, the momentum basis provides the most compact representation, consistent with the approach toward a single Fermi-sea configuration as \(U/t_{\rm leg}\rightarrow0\). For strong rung hopping, the bonding/antibonding basis becomes favorable because it diagonalizes the dominant local rung structure, whereas the leg-mode basis becomes advantageous when intra-leg hopping dominates. At stronger onsite interaction, the position basis becomes increasingly favorable as double occupation is suppressed. The corresponding basis-resolved compressibility landscapes for all four representations are provided in Sec.~\ref{sec:AdditionalCompressibility} and Fig.~\ref{fig:CompressibilitybyBasis12Site} of the Supplementary Information.

Importantly, selecting the most favorable one-body representation does not produce a uniformly sparse ground state. Figure~\ref{fig:12SiteCompressibility}(b) shows the minimum normalized subspace fraction, \(f_W\), after choosing the most compressible basis at each parameter point. A least-compressible region remains near intermediate interaction strength and approximately isotropic hopping, \(t_{\rm rung}\approx t_{\rm leg}\), where none of the physically motivated one-body representations is clearly adapted to the correlated ground state. At the representative point \(U/t_{\rm leg}=5\) and \(t_{\rm rung}/t_{\rm leg}=1\), approximately \(3.5\%\) of the fixed-sector Hilbert space is required to retain \(99\%\) of the ground-state weight for the 12-site system.

The persistence and scaling of this poorly compressible regime are further examined in Fig.~\ref{fig:12SiteCompressibility}(c,d). At half filling, the full fixed-\(S_z\) Hilbert-space dimension is \(D_{\rm full}(N)=\binom{N}{N/2}^{2}\), which has the Stirling asymptotic form \(D_{\rm full}(N)\sim [2/(\pi N)]4^{N}\). Motivated by this combinatorial scaling, the required subspace size at cumulative ground-state weight \(W\) is guessed to asymptotically follow \(k_W(N)\sim (B_W/N)q_W^{N}\), where \(B_W\) and \(q_W\) are obtained from fits to the computed system sizes. Correspondingly, the retained subspace fraction scales as \(k_W/D_{\rm full}\sim(\pi B_W/2)(q_W/4)^{N}\). Figure~\ref{fig:12SiteCompressibility}(c) compares the computed \(99\%\) subspace fractions for the position, bonding/antibonding, leg-mode, and momentum representations at \(U/t_{\rm leg}=5\) and \(t_{\rm rung}/t_{\rm leg}=1\), together with the Stirling-motivated fit for the most favorable representation. For this threshold, \(B_{0.99}=0.435\) and \(q_{0.99}=3.106\), with \(R^2=0.973\) in log space. Figure~\ref{fig:12SiteCompressibility}(d) shows the absolute subspace size \(k_W\) as a function of system size for several retained-weight thresholds from the same parameter space as Figure~\ref{fig:12SiteCompressibility}(c), \(U/t_{\rm leg}=5\) and \(t_{\rm rung}/t_{\rm leg}=1\), using results from even system sizes between 2 and 14 sites, with fitted curves extrapolated to larger ladders. Although the fitted growth factors satisfy \(q_W<4\), so that the required fraction of the full Hilbert space decreases with system size, the absolute subspace size \(k_W\) continues to grow exponentially. These results demonstrate that physically motivated one-body orbital rotations can substantially improve compressibility in limiting parameter regimes, but do not remove the rapidly growing basis-state requirement of sample-based diagonalization in the intermediate correlated regime.

\subsection{Statevector vs \newline Exact Diagonalization}

Unless otherwise stated, this section primarily focuses on the half-filled 10-site model with \(S_z=0\), intermediate interaction strength \(U/t=4\), and isotropic hopping \(t_{\rm rung}/t_{\rm leg}=t\). To separate the intrinsic compressibility of a chosen orbital basis from the ability of the SKQD circuit to identify important configurations, we construct an exact-diagonalization reference for each basis. The exact ground state is expanded in the corresponding occupation-number basis, and the configurations are ranked by their exact ground-state weights \(w_i=|\alpha_i|^2\), where \(|\Psi_0\rangle=\sum_i\alpha_i|\phi_i\rangle\). For a retained subspace of dimension \(k\), the \(k\) highest-weight configurations are selected to define 

\begin{equation}
\mathcal{S}_k
=
\operatorname{span}
\left\{
|\phi_1\rangle,\ldots,|\phi_k\rangle
\right\},
\end{equation}
where the configurations are ordered such that \(w_1\geq w_2\geq\cdots\). The Hamiltonian is projected onto \(\mathcal{S}_k\), and the lowest eigenvalue of the projected Hamiltonian is recalculated. Repeating this procedure with increasing \(k\) produces the solid reference curves in Fig.~\ref{fig:StatevectorParameterPerformance}.

The reference curves in Fig.~\ref{fig:StatevectorParameterPerformance} represent an idealized limit in which the configurations carrying the largest exact ground-state weight are known a priori. They therefore provide a best-case ground-state-weight benchmark for configuration selection within each orbital basis. The corresponding statevector curves instead use configurations identified from the measurement distributions generated by the SKQD circuits. The separation between the statevector and reference curves consequently measures the additional inefficiency introduced by the circuit sampling procedure beyond the intrinsic compressibility of the selected basis.

\subsubsection{Reference State Preparation}

The statevector calculations primarily use a momentum-space Fermi-sea reference state, which is expanded using particle-number-conserving \(XX+YY\) mixing layers before Trotter evolution as described in Sec.~\ref{sec:SKQDCircuitPreparation} of the Supplementary Information. These layers broaden the initially compact Fermi-sea configuration into a controlled superposition intended to increase overlap with correlated components of the ground state. Too little mixing may leave the reference state insufficiently expressive, whereas excessive mixing can distribute probability across configurations with comparatively small ground-state weight. The number of mixing layers therefore provides a tunable balance between enriching the reference state and maintaining a selective sampling distribution. The complete reference-state, orbital-rotation, and Krylov-circuit construction is summarized in Sec.~\ref{sec:SKQDCircuitPreparation} and Fig.~\ref{fig:SKQDCircuit} of the Supplementary Information. The dependence of the reconstructed subspace on this parameter is examined explicitly in Figs.~\ref{fig:StatevectorParameterPerformance}(c,g).

\subsubsection{Statevector Sampling Parameter Dependence}

\begin{figure*}[!t]
    \centering
    \includegraphics[width=\textwidth]{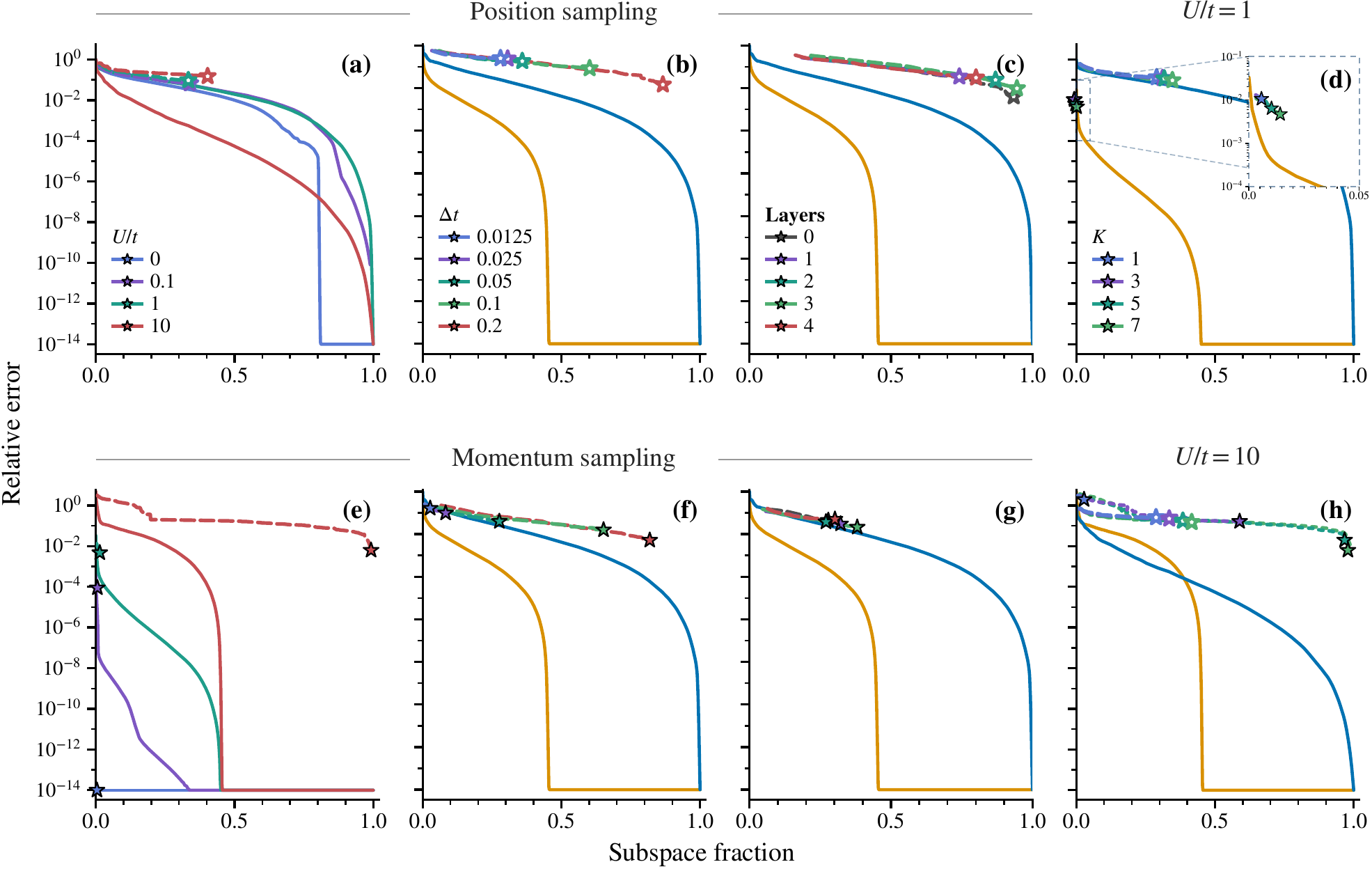}
    \caption{Statevector reconstruction performance as a function of retained subspace fraction for the half-filled 10-site Fermi-Hubbard ladder. Panels (a-c) and (e-g) show position- and momentum-basis sampling, respectively, while varying \(U/t\) (a,e), Trotter step size \(\Delta t\) (b,f), and the number of initial particle-number-conserving \(XX+YY\) mixing layers (c,g). Panels (d,h) compare both sampling bases as a function of maximum Trotter-step count \(K\) at \(U/t=1\) and \(10\), respectively. Solid curves denote exact-diagonalization references and dashed curves denote statevector-sampled reconstructions; stars mark the terminal reconstructed subspaces. Unless varied, \(K=7\), \(\Delta t=0.025\), and three \(XX+YY\) mixing layers are used. Sampling uses 200 samples per Krylov circuit for the layer sweep and 2,000 for the remaining studies. }
    \label{fig:StatevectorParameterPerformance}
\end{figure*}

We first examine how SKQD sampling changes with interaction strength. Figures~\ref{fig:StatevectorParameterPerformance}(a,e) compare position- and momentum-basis sampling, respectively, with their exact-diagonalization reference curves for the half-filled 10-site ladder over \(U/t=0\) to \(10\). Figure~\ref{fig:StatevectorParameterPerformance}(e) demonstrates the pronounced advantage of momentum-basis sampling in the weakly to intermediately interacting regime. Although the Trotter evolution itself is implemented in the position basis, the momentum-sampled statevector curves remain relatively close to their corresponding exact-diagonalization references for \(U/t\lesssim1\), indicating that the circuit preferentially identifies many of the configurations carrying the dominant ground-state weight. The position-sampled curves in Fig.~\ref{fig:StatevectorParameterPerformance}(a) similarly track their corresponding exact-diagonalization references at low interaction strength, while the position-basis reference curves themselves vary less strongly with \(U/t\) than their momentum-basis counterparts. This indicates that the intrinsic compressibility of the position representation is comparatively less sensitive to interaction strength over the range considered. As \(U/t\) increases, the momentum-basis advantage decreases as the interacting ground state becomes less compact in that representation. By \(U/t=10\), the reconstructed performance of momentum- and position-basis sampling becomes comparable.

We next examine the dependence of SKQD sampling on the Krylov dimension \(K+1\), where \(K\) denotes the maximum number of Trotter steps, as shown in Figs.~\ref{fig:StatevectorParameterPerformance}(d,h). At fixed \(\Delta t=0.025\), increasing \(K\) introduces additional time-evolved Krylov states and increases the maximum elapsed evolution time according to \(T_{\max}=K\Delta t\). Figures~\ref{fig:StatevectorParameterPerformance}(d) and \ref{fig:StatevectorParameterPerformance}(h) therefore characterize the combined effect of Krylov dimension and elapsed evolution time for the \(U/t=1\) and \(U/t=10\) systems, respectively. Increasing \(K\) initially enriches the sampled configuration space with basis states relevant to the ground state. At sufficiently large \(K\), however, the sampled distribution broadens and increasingly includes configurations carrying comparatively small ground-state weight. This behavior is particularly apparent in Fig.~\ref{fig:StatevectorParameterPerformance}(h), where the final reconstructed subspace spans approximately \(99.2\%\) of the fixed particle-number sector while retaining a relative energy error of approximately \(6\times10^{-3}\). Analysis of the omitted configurations shows that this residual error is dominated by a small subset of basis states that are important to the low-energy subspace, demonstrating that reconstruction quality can remain sensitive to the omission of even a small number of important configurations.

The dependence on Trotter step size is shown in Figs.~\ref{fig:StatevectorParameterPerformance}(b) and \ref{fig:StatevectorParameterPerformance}(f). These calculations use fixed \(K=7\) and three initial mixing layers while varying \(\Delta t\). Because the sampled Krylov states span \(m=0,\ldots,K\), varying \(\Delta t\) also changes the maximum evolution time, \(T_{\max}=7\Delta t\). Smaller \(\Delta t\) therefore corresponds both to more closely spaced Krylov states and to a shorter total evolution window. In this regime, the reconstructed configurations remain more selectively concentrated on basis states with large ground-state weight. As \(\Delta t\) increases, the time-evolved states explore a broader portion of the symmetry-allowed configuration space and the selectivity of the sampled subspace decreases. The similar behavior observed for \(\Delta t=0.1\) and \(0.2\) suggests that, once the sampled distribution becomes sufficiently broad, reconstruction is increasingly limited by the retained subspace rather than by access to additional configurations. Consequently, increasing either the evolution interval or the sampled configuration breadth does not necessarily improve the efficiency of the recovered subspace.

Figures~\ref{fig:StatevectorParameterPerformance}(c,g) show the dependence of sampling on the number of particle-number-conserving \(XX+YY\) mixing layers applied before Trotter evolution for the half-filled 10-site ladder at \(U/t=4\). Increasing the number of layers modestly broadens the initialized distribution in both sampling bases. Three layers provide a favorable compromise between expanding beyond the initial Fermi-sea reference and avoiding excessive population of low-importance configurations, whereas the four-layer construction produces a broader distribution without a corresponding improvement in reconstructed energy. The poorer position-basis performance is also consistent with the use of a momentum-space Fermi-sea reference state, which is naturally better adapted to momentum-basis sampling. A position-adapted reference state, such as a N\'eel-like or variationally optimized state, may therefore improve position-basis sampling, but a systematic comparison of reference-state constructions is beyond the scope of the present work.

Taken together, Fig.~\ref{fig:StatevectorParameterPerformance} shows that statevector SKQD performance is controlled jointly by the intrinsic compressibility of the selected orbital basis and by the sampling dynamics of the circuit. A favorable basis alone does not guarantee efficient reconstruction: reference-state preparation, Krylov depth, Trotter step size, and the resulting evolution window must collectively produce a measurement distribution concentrated on configurations with large ground-state weight. Conversely, excessive state spreading can substantially enlarge the sampled configuration space without producing a correspondingly more accurate reduced subspace. This distinction between intrinsic basis compressibility and circuit sampling efficiency provides the basis for interpreting the hardware results below.

\subsection{Quantum Hardware Results}

We next apply the insights obtained from the statevector simulations to larger Fermi-Hubbard ladders sampled on the square-topology \texttt{ibm\_miami} processor over interaction strengths ranging from \(U/t=0\) to \(10\). We first consider the half-filled 20-site ladder to characterize the effect of hardware noise on the sampled particle-number sector. The hardware circuits use the adapted FSWAP construction described in Sec.~\ref{sec:FSWAPNetwork} of the Supplementary Information to reduce routing and Jordan-Wigner parity-string overhead. Additional mapping and transpilation resource comparisons for the hardware SKQD circuits are provided in Sec.~\ref{sec:HardwareResources}. For each calculation on this model, \(K+1=8\) Krylov circuits are sampled, corresponding to \(m=0,\ldots,7\) and therefore up to seven Trotter steps in the deepest circuit, with \(50{,}000\) shots collected per circuit.

\begin{figure}[!t]
    \centering
    \includegraphics[width=\columnwidth]{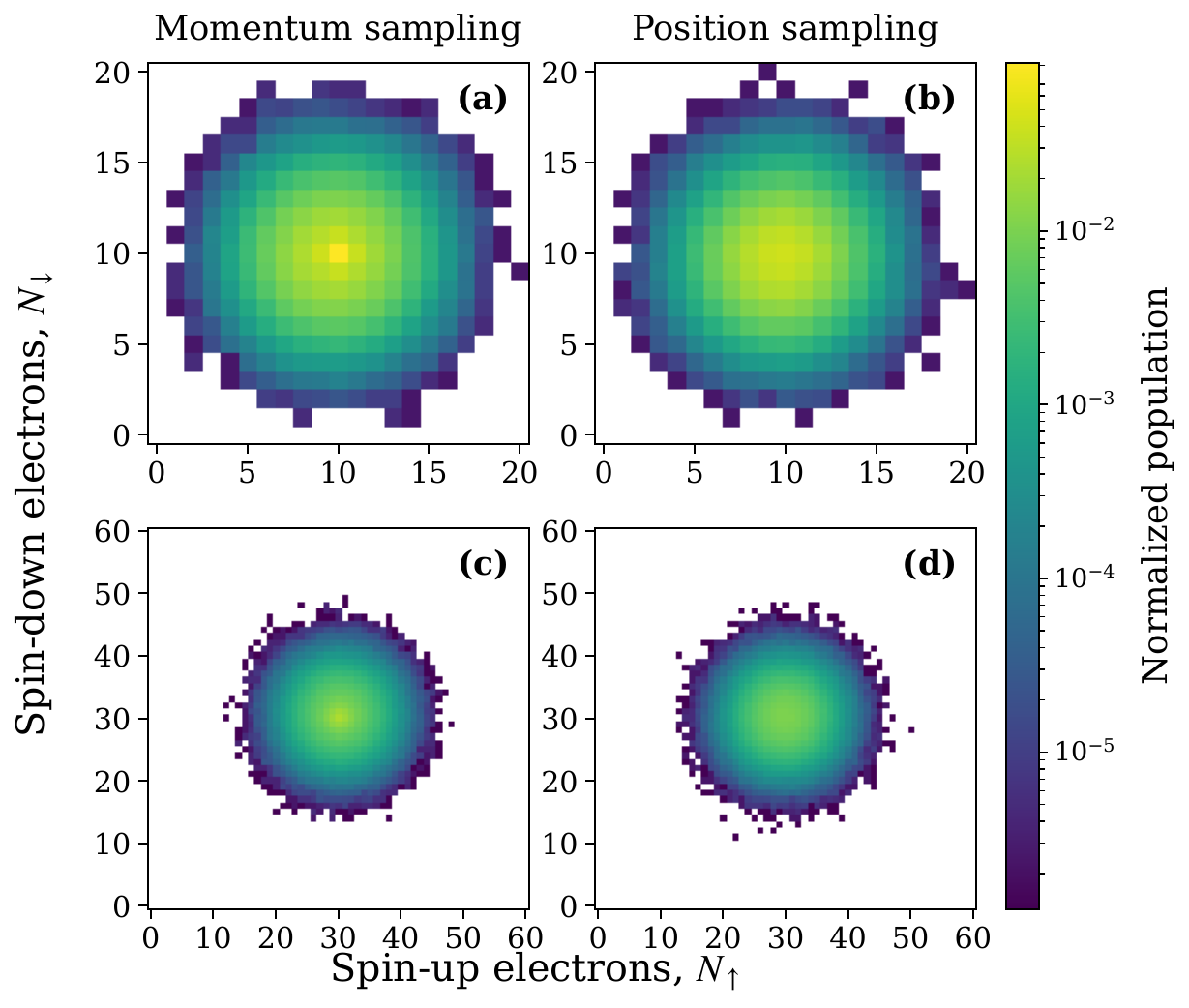}
    \caption{Particle-number sector distributions of hardware samples. The normalized sampled population is plotted as a function of spin-up electrons $N_{\uparrow}$ and spin-down electrons $N_{\downarrow}$ for the 20-site ladder using (a) momentum-basis and (b) position-basis sampling, and for the 60-site ladder using (c) momentum-basis and (d) position-basis sampling. The distributions are centered near the half-filled sectors \((N_{\uparrow},N_{\downarrow})=(10,10)\) and \((30,30)\) for the 20- and 60-site systems, respectively.}
    \label{fig:HardwareHammingWeight}
\end{figure}

Before configuration recovery, we examine the joint Hamming-weight distribution of the measured spin-up and spin-down sectors, as shown in Fig.~\ref{fig:HardwareHammingWeight} for the \(U/t=1.0\) sector. Here, configuration recovery refers to the SQD post-processing procedure used to recover symmetry-consistent occupation-number configurations from noisy measured bitstrings prior to construction of the reduced diagonalization subspace~\cite{robledo2025chemistry}. For an ideal circuit modeling the half-filled 20-site system, all sampled configurations would remain in the target sector \((N_{\uparrow},N_{\downarrow})=(10,10)\), where \(N_{\uparrow}\) and \(N_{\downarrow}\) correspond to the number of electrons in the spin-up and spin-down configurations. Hardware noise instead produces population in neighboring particle-number sectors, providing a direct measure of symmetry leakage in the raw measurement distribution. Configuration recovery is subsequently applied to the hardware samples before constructing the reduced Hamiltonian subspaces used for energy reconstruction.

Figures~\ref{fig:HardwareHammingWeight}(a,b) show the corresponding raw particle-number distributions for the 20-site ladder at \(U/t=1\). Although hardware noise spreads the samples over a broad range of \((N_{\uparrow},N_{\downarrow})\) sectors, the distribution remains centered near the target half-filled sector \((10,10)\). Momentum-basis sampling retains approximately \(9.3\%\) of the unprocessed samples directly in the target sector, compared with approximately \(4.2\%\) for position-basis sampling. To further characterize this spread, we define the minimum Hamming distance to the target particle-number sector as \[d_H^{\rm sector}=|N_{\uparrow}-10|+|N_{\downarrow}-10|,\] which corresponds to the minimum number of bit flips required to restore the correct spin-resolved particle numbers. The mean sector distance is approximately \(3.1\) for momentum-basis sampling and \(3.2\) for position-basis sampling. Thus, while only a fraction of the raw samples remain exactly in the target sector, the measured distributions retain a clear concentration around the intended half-filled particle numbers.

Similar behavior is observed for the 60-site ladder in Figs.~\ref{fig:HardwareHammingWeight}(c,d), where the target sector is \((N_{\uparrow},N_{\downarrow})=(30,30)\). Here, \(100{,}000\) shots are collected per Krylov circuit. As the system size increases, the fraction of samples remaining exactly in the target sector decreases to approximately \(2.3\%\) for momentum-basis sampling and \(1.0\%\) for position-basis sampling. The corresponding mean minimum distances from the target sector increase to approximately \(5.7\) and \(6.2\), respectively. Nevertheless, the distributions remain centered near \((30,30)\), indicating that the hardware samples continue to retain information about the intended particle-number sector even at the 60-site scale.

The systematically larger target-sector population obtained from momentum-basis sampling may be influenced by the momentum-space Fermi-sea reference used for both circuit constructions. In particular, the \(m=0\) momentum-basis circuit requires no Trotter evolution beyond preparation of the reference configuration, whereas obtaining position-basis samples from the same reference requires an additional orbital rotation. For larger Krylov indices, however, circuit depth depends on both the Trotter evolution and the orbital rotations required for the selected measurement basis, so the observed basis dependence cannot be attributed to circuit depth alone. Rather, it reflects a combination of reference-state preparation, basis-rotation overhead, accumulated hardware noise, and the resulting sampling distribution.

\begin{figure*}[!t]
    \centering
    \includegraphics[width=.92\textwidth]{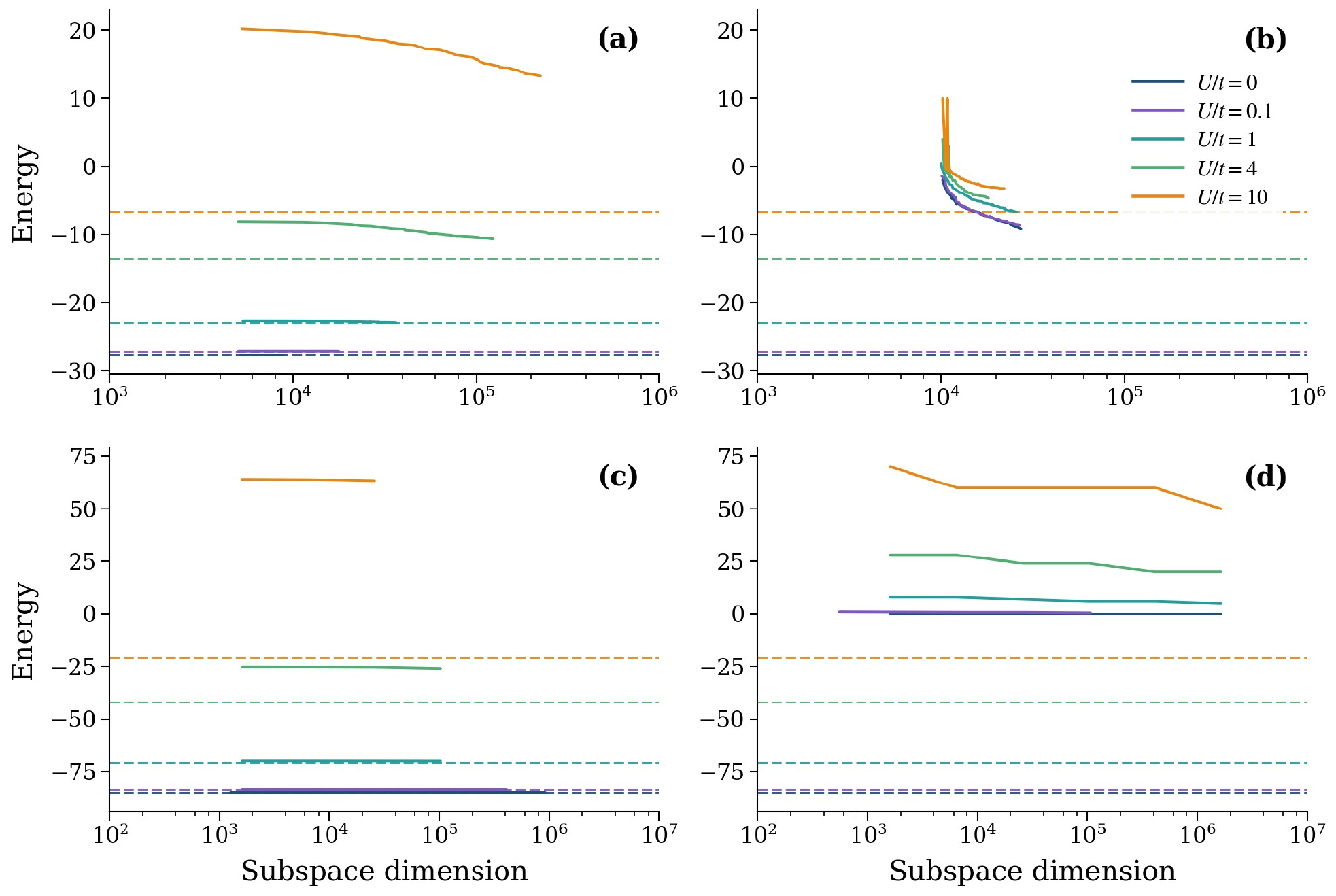}
    \caption{Hardware-reconstructed energy as a function of retained subspace dimension for the 20-site (a,b) and 60-site (c,d) Fermi-Hubbard ladders sampled on \texttt{ibm\_miami}. Panels (a,c) show momentum-basis sampling and panels (b,d) show position-basis sampling for interaction strengths \(U/t=0\), \(0.1\), \(1\), \(4\), and \(10\). Solid curves show energies reconstructed using the SKQD-Fulqrum workflow, while horizontal dashed lines denote the corresponding DMRG ground-state reference energies.}
    \label{fig:HardwareEnergy}
\end{figure*}

Despite substantial symmetry leakage in the raw bitstrings, the concentration of the measured distributions around the correct half-filled sectors provides a meaningful starting point for configuration recovery.   We apply the self-consistent configuration-recovery procedure introduced for SQD~\cite{robledo2025chemistry}, using a Fulqrum-based workflow~\cite{nation2026generalized,qiskitcommunity2026fulqrum} following the structure implemented in \texttt{qiskit-addon-sqd}~\cite{qiskitaddonsqd2024}. The procedure uses the noisy hardware distribution together with the known particle-number constraints to construct symmetry-valid configurations, which are subsequently used to define the reduced Hamiltonian subspaces examined in Fig.~\ref{fig:HardwareEnergy}.

Figure~\ref{fig:HardwareEnergy} compares the reconstructed hardware energies with the corresponding DMRG reference energies as the retained configuration subspace is increased. Convergence of the DMRG ground-state energies used as references is documented in Sec.~\ref{sec:DMRGReference} and Fig.~\ref{fig:DMRGTruncationErr} of the Supplementary Information. Consistent with the statevector results, momentum-basis sampling performs most favorably in the weakly interacting regime. For both the 20-site and 60-site ladders in Figs.~\ref{fig:HardwareEnergy}(a,c), the reconstructed energies for \(U/t\leq1\) approach the corresponding DMRG references substantially more closely than those obtained at stronger interaction. As \(U/t\) increases toward \(4\) and \(10\), the reconstructed energies remain appreciably separated from the reference values over the accessible subspace dimensions, indicating that the present hardware-sampling and recovery procedure does not identify a sufficiently effective reduced subspace in this more strongly correlated regime.

The position-basis results in Figs.~\ref{fig:HardwareEnergy}(b,d) display a weaker and more size-dependent relationship with interaction strength than the momentum-basis results. For the 20-site ladder, the relative performance of position-basis sampling improves as \(U/t\) increases, and at \(U/t=10\) the reconstructed energy approaches the DMRG reference more closely than the corresponding momentum-basis result. For the 60-site ladder, however, the weakly interacting position-basis reconstructions remain far from their reference energies even as the retained subspace becomes large. The stronger-interaction calculations recover more substantial energy lowering, but still remain well separated from the corresponding DMRG references. These results show that the basis dependence predicted from intrinsic ground-state compressibility can be strongly modified by system size, hardware sampling, and the subsequent configuration-recovery procedure.

A particularly clear limitation of position-basis sampling appears for the 60-site noninteracting system in Fig.~\ref{fig:HardwareEnergy}(d). At \(U/t=0\), the Fermi-Hubbard Hamiltonian in the position basis contains no diagonal interaction contribution, and its nonzero matrix elements arise entirely from hopping between occupation-number configurations connected by a single allowed fermionic hop. The quality of a recovered subspace therefore depends not only on which configurations are sampled individually, but also on whether those configurations are connected to one another by the Hamiltonian.

This connectivity can be understood from a simple four-site example. Consider the occupation-number configurations \(\ket{1001}_{\mathrm{site},\uparrow}\otimes\ket{1010}_{\mathrm{site},\downarrow}\) and \(\ket{0101}_{\mathrm{site},\uparrow}\otimes\ket{1010}_{\mathrm{site},\downarrow}\), which are connected by moving the corresponding spin-up particle from the first site to the second site. Although the bitstrings differ at two bit positions, they are adjacent in the determinant-connectivity graph because they are related by a single physical hopping process. Consequently, \(\langle01011010|\hat H_t|10011010\rangle\neq0\), producing an off-diagonal matrix element proportional to \(-t\) in the Hamiltonian projected onto a subspace containing both configurations.

To quantify this property, we define a graph over the retained configuration set \(\mathcal{S}_k\), following determinant-space Hamiltonian connectivity used in FCIQMC and more recent active-sampling SQD approaches~\cite{booth2009fermion,miura2026active}. Each sampled determinant is represented as a vertex, and an edge is placed between configurations \(i\) and \(j\) whenever \(\langle\phi_i|\hat H_t|\phi_j\rangle\neq0\). The number of hopping-connected determinant pairs is therefore

\begin{equation}
N_{\rm edge}(k)
=
\frac{1}{2}
\sum_{\substack{i\neq j\\ i,j\in\mathcal{S}_k}}
\mathbf{1}
\left[
\langle\phi_i|\hat H_t|\phi_j\rangle\neq0
\right],
\end{equation}
while the corresponding mean connectivity is

\begin{equation}
\bar d(k)
=
\frac{2N_{\rm edge}(k)}{k}.
\end{equation}
These quantities distinguish the number of configurations retained from the amount of hopping connectivity \(\hat H_t\) supported within the retained subspace.

This distinction is especially relevant to the \(U/t=0\) position-basis result in Fig.~\ref{fig:HardwareEnergy}(d). If the recovered configuration set contains no hopping-connected pairs, then \(N_{\rm edge}=0\). Because the position-basis Hamiltonian also has zero diagonal contribution at \(U=0\), the Hamiltonian projected into such a subspace is identically zero and its minimum eigenvalue remains \(E=0\), regardless of how many mutually disconnected configurations are retained. Increasing the subspace dimension therefore does not necessarily improve the reconstruction if the sampled states do not simultaneously acquire Hamiltonian connectivity.

As the many-body configuration space grows combinatorially with system size, this connectivity requirement becomes increasingly important. Two determinants connected by a single hopping process occupy nearby positions in the determinant-connectivity graph, but sampling a large number of configurations independently does not guarantee that many such connected pairs will be recovered. A subspace may therefore contain a large number of individually sampled configurations while remaining highly disconnected under the hopping Hamiltonian, limiting its ability to reproduce the energy-lowering kinetic contribution.

The \(U/t=0.1\) position-basis result in Fig.~\ref{fig:HardwareEnergy}(d) exhibits a closely related limitation. Although the weak onsite interaction introduces nonzero diagonal contributions to the projected Hamiltonian, the reconstructed energy remains near zero over the accessible subspace dimensions and far above the corresponding DMRG ground-state energy. Increasing the number of retained configurations, therefore does not by itself recover the large negative kinetic-energy contribution expected for the weakly interacting ground state. This behavior is consistent with a recovered configuration set that contains many symmetry-valid determinants but insufficient hopping connectivity between them.

More generally, these results show that successful SQD requires more than identifying individually important configurations. The recovered states must collectively form a sufficiently connected subspace under the target Hamiltonian, such that the hopping processes responsible for lowering the many-body energy are represented within the projected matrix. Subspace dimension and Hamiltonian connectivity should therefore be viewed as complementary measures of the usefulness of a recovered configuration set. A controlled comparison in which momentum- and position-derived hardware samples are evaluated using the same position-basis reconstruction Hamiltonian is provided in Sec.~\ref{sec:SamplingDependence} of the Supplementary Information, further demonstrating that reconstruction quality depends on the sampled configuration set itself rather than on subspace dimension or Hamiltonian representation alone.

\FloatBarrier
\section{Conclusion}

In this work, we investigated the consequences of basis-dependent sampling for the Fermi-Hubbard ladder using physically motivated one-body orbital rotations. The resulting ground-state compressibility exhibits a strong dependence on the onsite interaction \(U\) and the anisotropic hopping amplitudes \(t_{\rm rung}\) and \(t_{\rm leg}\), with different orbital representations becoming favorable in distinct physical regimes. Statevector and quantum-hardware SKQD calculations show that momentum-basis sampling can reproduce reference ground-state energies obtained from exact diagonalization and density matrix renormalization group with comparatively high accuracy in the weakly to intermediately interacting regime. As the interaction strength increases, the intrinsic compressibility advantage of the momentum representation decreases and the position representation becomes increasingly competitive. However, near the intermediate, approximately isotropic correlated regime, the number of basis states required to retain a large fraction of the ground-state weight is consistent with exponential growth over the computed system sizes. At the representative intermediate-coupling point \(U/t_{\rm leg}=5\) and \(t_{\rm rung}/t_{\rm leg}=1\), approximately \(3.5\%\) of the fixed-sector Hilbert space is required to retain \(99\%\) of the ground-state weight for the 12-site ladder. Across the computed system sizes, the corresponding absolute subspace requirement is described by \(k_{0.99}(N)\sim (0.435/N)(3.106)^N\), indicating that basis selection can reduce, but does not eliminate, the increasing classical subspace cost imposed by the intrinsic compressibility of the model.

The statevector SKQD calculations further suggest that intrinsic compressibility alone does not determine sampling efficiency. For the 10-site ladder at \(U/t=1\), momentum-basis sampling with \(K=7\) reaches a relative energy error of approximately \(4.7\times10^{-3}\) using only \(1.4\%\) of the fixed particle-number sector, whereas position-basis sampling reaches approximately \(9.5\times10^{-2}\) relative error while retaining about \(33\%\) of the sector under the same Krylov construction. At stronger interaction, the relationship between retained subspace size and reconstruction accuracy becomes less direct. At \(U/t=10\), for example, the momentum-sampled reconstruction spans approximately \(99.2\%\) of the fixed particle-number sector while retaining a relative energy error of approximately \(6.3\times10^{-3}\). 

These sampling constraints become increasingly important on quantum hardware. For the 20-site and 60-site ladders sampled on the square-topology \texttt{ibm\_miami} processor, momentum-basis reconstruction at \(U/t=1\) agrees with the corresponding DMRG reference energies to within approximately \(2.9\times10^{-3}\) and \(1.2\times10^{-2}\) in relative energy, respectively. At the same time, the fraction of raw momentum-basis samples remaining directly in the target particle-number sector decreases from approximately \(9.3\%\) for the 20-site system to \(2.3\%\) for the 60-site system. Configuration recovery can restore symmetry-consistent configurations, but the resulting subspace must also preserve the Hamiltonian connectivity required to reproduce the energy-lowering hopping processes. The hardware results therefore show that large recovered subspaces can remain ineffective when they contain poorly connected or low-importance configurations, emphasizing that subspace dimension alone is not a sufficient measure of reconstruction quality.

Taken together, these results show that the effectiveness of SKQD for strongly correlated fermionic systems is controlled jointly by the intrinsic compressibility of the chosen representation, the configuration distribution generated by the quantum circuit, and the structure of the classically reconstructed subspace. The correlation regime of the underlying model can therefore provide a useful guide for selecting sampling representations, but basis optimization alone does not remove the approximated exponential growth of the relevant many-body configuration space or guarantee efficient quantum sampling. Future improvements may come from variationally optimized orbital rotations, basis-adapted reference states, and recovery procedures that explicitly account for the importance and Hamiltonian connectivity of sampled configurations. Extending sample-based diagonalization to larger and more strongly correlated systems will therefore require coordinated optimization of the sampled basis, quantum circuit, and classical reconstruction procedure.

\section{Acknowledgments}

This work was supported by IBM through the IBM Rensselaer Future of Computing Research Collaboration. We thank the Quantum Algorithm Engineering team at IBM for valuable discussions and technical support scaling up quantum workflows. Jian Shi thanks the Simons Foundation for the Pivot Fellowship. We would like to express our sincere gratitude to Nate Earnest Noble, and Tiffany Ly from IBM for their valuable suggestions and insightful discussions throughout this work. We also thank Michael Sofka from FOCI RPI for his technical help in using the quantum hardware. We also acknowledge the use of Codex (OpenAI) for assistance in developing, revising, and debugging selected data analysis and plotting scripts used in this work. The authors reviewed and validated all AI-assisted code before use.

\section{Data and Code Statement}

Data and code for this work are available from
the corresponding author upon request.

\balance
\bibliographystyle{unsrt}
\bibliography{SKQD}

\clearpage
\nobalance

\section*{Supplementary Information}

\setcounter{section}{0}
\renewcommand{\thesection}{S\arabic{section}}

\setcounter{figure}{0}
\renewcommand{\thefigure}{S\arabic{figure}}

\setcounter{equation}{0}
\renewcommand{\theequation}{S\arabic{equation}}

\section{Two-Site \newline Basis-Rotation Analysis}
\label{sec:TwoSiteBasisAnalysis}

Here we consider the half-filled two-site Fermi-Hubbard model as a minimal system for illustrating how basis transformations modify the structure of the Hamiltonian. This two-site problem corresponds directly to a single rung of the ladder geometry and therefore provides a useful setting for comparing the one-body orbital rotations employed throughout this work with more general correlated many-body transformations. In particular, it illustrates both the potential simplification obtained by adapting the basis to a dominant term in the Hamiltonian and the additional complexity that can be transferred to the remaining interactions. In the fixed-\(S_z=0\) sector, the position-basis configurations are

\begin{equation}
\begin{aligned}
\ket{D_1} &= \ket{\uparrow\downarrow,0},\\
\ket{C_1} &= \ket{\uparrow,\downarrow},\\
\ket{C_2} &= \ket{\downarrow,\uparrow},\\
\ket{D_2} &= \ket{0,\uparrow\downarrow}.
\end{aligned}
\end{equation}
The corresponding two-site Fermi-Hubbard Hamiltonian is

\begin{equation}
\begin{aligned}
\label{eq:2siteHamiltonian}
\hat H= -t
[\left(
\hat{c}^{\dagger}_{0\downarrow}\hat{c}_{1\downarrow}
+
\hat{c}^{\dagger}_{1\downarrow}\hat{c}_{0\downarrow}
\right)
+
\left(
\hat{c}^{\dagger}_{0\uparrow}\hat{c}_{1\uparrow}
+
\hat{c}^{\dagger}_{1\uparrow}\hat{c}_{0\uparrow}
\right)
]
\\
+U
\left(
\hat{n}_{0\uparrow}\hat{n}_{0\downarrow}
+
\hat{n}_{1\uparrow}\hat{n}_{1\downarrow}
\right).
\end{aligned}
\end{equation}
Here, \(\ket{D_1}\) and \(\ket{D_2}\) denote configurations containing a doubly occupied site, while \(\ket{C_1}\) and \(\ket{C_2}\) denote the two covalent configurations with one fermion on each site. The operator \(\hat{c}_{i\sigma}\) annihilates a fermion with spin \(\sigma\) on site \(i\), \(t\) is the nearest-neighbor hopping amplitude, \(U\) is the onsite interaction strength, and \(\hat{n}_{i\sigma}=\hat{c}^{\dagger}_{i\sigma}\hat{c}_{i\sigma}\) is the corresponding number operator. In the ordered basis \(\left[\ket{D_1},\ket{C_1},\ket{C_2},\ket{D_2}\right]\), the Hamiltonian takes the matrix form

\begin{equation}
H =
\begin{pmatrix}
U & -t & -t & 0\\
-t & 0 & 0 & -t\\
-t & 0 & 0 & -t\\
0 & -t & -t & U
\end{pmatrix}.
\end{equation}

We next consider two transformations of this Hamiltonian. The first is the one-body bonding/antibonding orbital rotation used elsewhere in this work, which acts directly on the single-particle fermionic modes. The second is a correlated transformation of the many-body configuration basis, which illustrates how additional compression can be obtained when the transformation is not restricted to one-body orbital rotations.

\subsection{One-Body Bonding/Antibonding Rotation}
\label{sec:BARotationAnalysis}

We first apply the one-body bonding/antibonding transformation defined in Eq.~\ref{eq:bAbRotation} to the two-site Hamiltonian in Eq.~\ref{eq:2siteHamiltonian}. Symmetric and antisymmetric combinations of the two-chain orbitals provide the conventional bonding and antibonding channels used in treatments of coupled Hubbard chains~\cite{balents1996weak,lin1997nchain}. This transformation diagonalizes the single-particle hopping term by introducing the bonding and antibonding orbitals, denoted by \(+\) and \(-\), respectively. In this basis, the Hamiltonian becomes

\begin{equation}
\begin{aligned}
\label{eq:bAbsiteHamiltonian}
\hat H= -t
\left[
\left(
\hat{n}_{+\uparrow}
-
\hat{n}_{-\uparrow}
\right)
+
\left(
\hat{n}_{+\downarrow}
-
\hat{n}_{-\downarrow}
\right)
\right]
\\
+\frac{U}{2}
\left[
\left(
\hat{n}_{+\uparrow}
+
\hat{n}_{-\uparrow}
\right)
\left(
\hat{n}_{+\downarrow}
+
\hat{n}_{-\downarrow}
\right)
\right.
\\
\left.
+
\left(
\hat{c}^{\dagger}_{+\uparrow}\hat{c}_{-\uparrow}
+
\hat{c}^{\dagger}_{-\uparrow}\hat{c}_{+\uparrow}
\right)
\left(
\hat{c}^{\dagger}_{+\downarrow}\hat{c}_{-\downarrow}
+
\hat{c}^{\dagger}_{-\downarrow}\hat{c}_{+\downarrow}
\right)
\right].
\end{aligned}
\end{equation}
The \(+\) and \(-\) labels denote occupation of the bonding and antibonding single-particle orbitals rather than the original physical sites. The corresponding fixed-\(S_z=0\), two-particle occupation basis is therefore

\begin{equation}
\mathcal{B}_{\rm BA}^{(2)}
=
\left[
\ket{\uparrow\downarrow,0}_{+,-},
\ket{\uparrow,\downarrow}_{+,-},
\ket{\downarrow,\uparrow}_{+,-},
\ket{0,\uparrow\downarrow}_{+,-}
\right],
\end{equation}
where the two entries now denote the bonding and antibonding orbitals. These states should not be interpreted as one-to-one images of the site-basis determinants; rather, they form the occupation-number basis associated with the rotated single-particle orbitals. In this ordered basis, the Hamiltonian takes the matrix form

\begin{equation}
H =
\begin{pmatrix}
\frac{U}{2}-2t & 0 & 0 & \frac{U}{2}\\
0 & \frac{U}{2} & \frac{U}{2} & 0\\
0 & \frac{U}{2} & \frac{U}{2} & 0\\
\frac{U}{2} & 0 & 0 & \frac{U}{2}+2t
\end{pmatrix}.
\end{equation}
The kinetic contribution is diagonal in this representation, but the onsite interaction is no longer diagonal. Instead, the transformed interaction generates off-diagonal couplings between bonding and antibonding configurations, including pair-transfer and spin-exchange-like processes, as expected under a general one-body orbital transformation~\cite{helgaker2000orbital}.

This distinction is important for quantum-circuit implementation. In the site basis, each hopping term in Eq.~\ref{eq:2siteHamiltonian} is a quadratic fermionic operator of the form
\(\hat{c}^{\dagger}_p\hat{c}_q+\hat{c}^{\dagger}_q\hat{c}_p\), which maps under the Jordan-Wigner transformation to an \(XX+YY\)-type interaction, together with the required parity string when the two modes are not adjacent in the Jordan-Wigner ordering. The onsite term \(U\hat{n}_{i\uparrow}\hat{n}_{i\downarrow}\) is diagonal in the occupation-number basis and can be implemented using controlled-phase interactions. In the bonding/antibonding basis, by contrast, the hopping contribution reduces to single-orbital phase evolution, while the transformed interaction contains off-diagonal quartic fermionic terms that generally produce more complicated Pauli-string rotations after Jordan-Wigner transformation~\cite{tranter2018comparison}. Consequently, a basis that provides a more compact ground-state representation does not necessarily provide a simpler basis in which to implement the Hamiltonian dynamics.

\subsection{Extension to the \newline Two-Leg Ladder}

The same one-body basis transformations can be extended directly from the two-site problem to the full two-leg Fermi-Hubbard ladder~\cite{noack1994correlations}. Their effect can be understood by separating the kinetic Hamiltonian into leg- and rung-hopping contributions. For a ladder containing \(N_{\rm rung}\) sites along each leg,

\begin{equation}
\begin{aligned}
\hat{H}_{t}
=
-t_{\rm leg}
\sum_{r=1}^{N_{\rm rung}-1}
\sum_{\ell=0}^{1}
\sum_{\sigma}
\left( \hat{c}_{r,\ell,\sigma}^{\dagger} \hat{c}_{r+1,\ell,\sigma}
+\hat{c}_{r+1,\ell,\sigma}^{\dagger} \hat{c}_{r,\ell,\sigma}
\right)
\\
-t_{\rm rung}
\sum_{r=1}^{N_{\rm rung}}
\sum_{\sigma}
\left( \hat{c}_{r,0,\sigma}^{\dagger} \hat{c}_{r,1,\sigma}
+\hat{c}_{r,1,\sigma}^{\dagger} \hat{c}_{r,0,\sigma}
\right).
\end{aligned}
\end{equation}
The bonding/antibonding basis applies the two-site rotation independently to each rung,

\begin{equation}
\hat{b}_{r,\pm,\sigma}
=
\frac{1}{\sqrt{2}}
\left(
\hat{c}_{r,0,\sigma}
\pm
\hat{c}_{r,1,\sigma}
\right).
\end{equation}
Under this transformation, the kinetic contribution becomes

\begin{equation}
\begin{aligned}
\hat{H}_{t}^{\rm BA}
={}&
-t_{\rm leg}
\sum_{r=1}^{N_{\rm rung}-1}
\sum_{\sigma}
\sum_{\nu=\pm}
\left( \hat{b}_{r,\nu,\sigma}^{\dagger} \hat{b}_{r+1,\nu,\sigma}
+\hat{b}_{r+1,\nu,\sigma}^{\dagger}
\hat{b}_{r,\nu,\sigma}
\right)
\\
&-
t_{\rm rung}
\sum_{r=1}^{N_{\rm rung}}
\sum_{\sigma}
\left(
\hat{n}_{r,+,\sigma}
-
\hat{n}_{r,-,\sigma}
\right).
\end{aligned}
\end{equation}
The rung-hopping contribution is therefore diagonalized locally on every rung, while the leg hopping remains a nearest-neighbor hopping process within the bonding and antibonding channels independently. This structure explains why the bonding/antibonding representation is naturally adapted to the regime \(t_{\rm rung}\gg t_{\rm leg}\), consistent with the enhanced compressibility observed in Fig.~\ref{fig:12SiteCompressibility}(a,b).

As in the two-site example, the simplification of the kinetic contribution is accompanied by additional structure in the transformed interaction. Defining
\begin{equation}
\hat{X}_{r,\sigma}
=
\hat{b}_{r,+,\sigma}^{\dagger}
\hat{b}_{r,-,\sigma}
+
\hat{b}_{r,-,\sigma}^{\dagger}
\hat{b}_{r,+,\sigma},
\end{equation}
the onsite interaction on the full ladder becomes

\begin{equation}
\hat{H}_{U}^{\rm BA}
=
\frac{U}{2}
\sum_{r=1}^{N_{\rm rung}}
\left[
\left(
\hat{n}_{r,+,\uparrow}
+
\hat{n}_{r,-,\uparrow}
\right)
\left(
\hat{n}_{r,+,\downarrow}
+
\hat{n}_{r,-,\downarrow}
\right)
+
\hat{X}_{r,\uparrow}
\hat{X}_{r,\downarrow}
\right].
\end{equation}
Thus, the local rung kinetic energy becomes diagonal, but the original density-density interaction generates off-diagonal couplings between bonding and antibonding orbitals, directly generalizing the two-site result above.

A complementary transformation is obtained by diagonalizing the one-body hopping independently along each leg. For open boundary conditions, the resulting single-particle eigenstates are standing-wave modes~\cite{lin1997nchain}, and the leg-mode operators may be written as

\begin{equation}
\hat{d}_{\ell m\sigma}
=
\sum_{r=1}^{N_{\rm rung}}
R^{\rm leg}_{rm}
\hat{c}_{r,\ell,\sigma},
\end{equation}
where
\begin{equation}
R^{\rm leg}_{rm}
=
\sqrt{\frac{2}{N_{\rm rung}+1}}
\sin
\left(
\frac{\pi r m}{N_{\rm rung}+1}
\right),
\qquad
m=1,\ldots,N_{\rm rung}.
\end{equation}
The hopping along the two legs is then diagonal,

\begin{equation}
\hat{H}_{\rm leg}
=
\sum_{\ell=0}^{1}
\sum_{m=1}^{N_{\rm rung}}
\sum_{\sigma}
\varepsilon_m
\hat{d}_{\ell m\sigma}^{\dagger}
\hat{d}_{\ell m\sigma},
\end{equation}
with

\begin{equation}
\varepsilon_m
=
-2t_{\rm leg}
\cos
\left(
\frac{\pi m}{N_{\rm rung}+1}
\right).
\end{equation}

Because the same orthogonal transformation is applied to both legs, the rung hopping is mode-diagonal and couples only equal-\(m\) orbitals on opposite legs,

\begin{equation}
\hat{H}_{\rm rung}
=
-t_{\rm rung}
\sum_{m=1}^{N_{\rm rung}}
\sum_{\sigma}
\left(
\hat{d}_{0m\sigma}^{\dagger}
\hat{d}_{1m\sigma}
+\hat{d}_{1m\sigma}^{\dagger}
\hat{d}_{0m\sigma}
\right).
\end{equation}
The onsite interaction, however, becomes nonlocal in the leg-mode indices,

\begin{equation}
\hat{H}_{U}^{\rm leg}
=
U
\sum_{\ell=0}^{1}
\sum_{mnpq}
\Gamma_{mnpq}
\hat{d}_{\ell m\uparrow}^{\dagger}
\hat{d}_{\ell n\downarrow}^{\dagger}
\hat{d}_{\ell p\downarrow}
\hat{d}_{\ell q\uparrow},
\end{equation}
where

\begin{equation}
\Gamma_{mnpq}
=
\sum_{r=1}^{N_{\rm rung}}
R^{\rm leg}_{rm}
R^{\rm leg}_{rn}
R^{\rm leg}_{rp}
R^{\rm leg}_{rq}.
\end{equation}
The leg-mode representation therefore provides the complementary simplification to the bonding/antibonding basis: it diagonalizes the dominant leg-hopping structure when \(t_{\rm leg}\gg t_{\rm rung}\), while transferring additional complexity into the interaction term. The corresponding improvement in ground-state compressibility in the leg-dominated regime is consistent with the behavior shown in Fig.~\ref{fig:12SiteCompressibility}(a,b).

These transformations illustrate that an orbital basis need not diagonalize the complete one-body Hamiltonian to provide a more compact representation of the interacting ground state. Instead, physically motivated rotations may selectively diagonalize the one-body substructures associated with the dominant geometry of the ladder. The bonding/antibonding and leg-mode bases are adapted to dominant rung and leg hopping, respectively, while the momentum basis considered in the following subsection diagonalizes the complete one-body hopping matrix.

\subsection{Relation to the Momentum Basis}

The bonding/antibonding construction can also be understood as the two-site limit of the momentum orbital transformation used in SKQD~\cite{yu2025quantum}. For the two-site system, the one-body hopping matrix is

\begin{equation}
h =
\begin{pmatrix}
0 & -t\\
-t & 0
\end{pmatrix}.
\end{equation}
Diagonalizing \(h\) gives the normalized eigenvectors
\[
\frac{1}{\sqrt{2}}
\begin{pmatrix}
1\\
1
\end{pmatrix},
\qquad
\frac{1}{\sqrt{2}}
\begin{pmatrix}
1\\
-1
\end{pmatrix},
\]
with eigenvalues \(-t\) and \(+t\), respectively. These eigenvectors are precisely the bonding and antibonding single-particle orbitals, so diagonalizing the complete one-body hopping matrix of the isolated two-site system reproduces the transformation in Eq.~\ref{eq:bAbRotation}. The bonding/antibonding and momentum bases are therefore equivalent for this minimal dimer.

This equivalence does not persist for larger Fermi-Hubbard ladders. The momentum basis diagonalizes the complete one-body hopping matrix, including both leg and rung hopping, whereas the bonding/antibonding basis diagonalizes only the local two-site hopping structure on each rung while retaining the real-space description along the ladder direction. Consequently, the two transformations generate distinct orbital representations and corresponding many-body Hamiltonian structures for larger systems, as illustrated by Figs.~\ref{fig:FHMatrixHamiltonian}(b) and \ref{fig:FHMatrixHamiltonian}(c).

\subsection{Correlated Many-Body Rotation}

We next consider a transformation that acts directly on the many-body configuration basis rather than on the underlying single-particle orbitals. Starting from the four fixed-\(S_z=0\) configurations of the two-site model, we define the symmetric and antisymmetric combinations

\begin{equation}
\begin{aligned}
\ket{D_+} &= \frac{1}{\sqrt{2}}
\left(\ket{D_1}+\ket{D_2}\right)
= \frac{1}{\sqrt{2}}
\left(\ket{\uparrow\downarrow,0}
+\ket{0,\uparrow\downarrow}\right),\\
\ket{S_0} &= \frac{1}{\sqrt{2}}
\left(\ket{C_1}+\ket{C_2}\right)
= \frac{1}{\sqrt{2}}
\left(\ket{\uparrow,\downarrow}
+\ket{\downarrow,\uparrow}\right),\\
\ket{T_0} &= \frac{1}{\sqrt{2}}
\left(\ket{C_1}-\ket{C_2}\right)
= \frac{1}{\sqrt{2}}
\left(\ket{\uparrow,\downarrow}
-\ket{\downarrow,\uparrow}\right),\\
\ket{D_-} &= \frac{1}{\sqrt{2}}
\left(\ket{D_1}-\ket{D_2}\right)
= \frac{1}{\sqrt{2}}
\left(\ket{\uparrow\downarrow,0}
-\ket{0,\uparrow\downarrow}\right).
\end{aligned}
\end{equation}
Here, \(\ket{D_+}\) and \(\ket{D_-}\) denote symmetric and antisymmetric doublon combinations, while \(\ket{S_0}\) and \(\ket{T_0}\) denote the corresponding covalent combinations. With the fermionic phase convention used in Eq.~\ref{eq:2siteHamiltonian}, \(\ket{S_0}\) belongs to the active ground-state block, whereas \(\ket{T_0}\) decouples from it.

In the ordered basis
\(\left[\ket{D_+},\ket{S_0},\ket{T_0},\ket{D_-}\right]\),
the Hamiltonian becomes

\begin{equation}
H =
\begin{pmatrix}
U & -2t & 0 & 0\\
-2t & 0 & 0 & 0\\
0 & 0 & 0 & 0\\
0 & 0 & 0 & U
\end{pmatrix}.
\end{equation}
The state \(\ket{T_0}\) is therefore decoupled from the interacting ground-state sector, while \(\ket{D_-}\) is an eigenstate with energy \(U\). The ground-state problem is consequently restricted to the two-dimensional subspace
\(\mathrm{span}\{\ket{D_+},\ket{S_0}\}\), within which

\begin{equation}
H_{\rm active}=
\begin{pmatrix}
U & -2t\\
-2t & 0
\end{pmatrix}.
\end{equation}
The corresponding ground-state energy is

\begin{equation}
E_0=
\frac{U-\sqrt{U^2+16t^2}}{2}.
\end{equation}
Thus, while the interacting ground state generally contains contributions from all four site-basis configurations, it is supported by only two configurations in this correlated basis. The two-site example therefore illustrates how a many-body transformation adapted to the correlation structure of the Hamiltonian can provide additional compression beyond a simple orbital rotation.

Unlike the one-body transformation \(R\) introduced in Eq.~\ref{eq:oneBodyRotation}, this change of basis acts directly on two-particle configurations. Defining the original pair-configuration creation operators as

\begin{equation}
\label{eq:sitePairOperators}
\begin{aligned}
\hat{P}_{D_1}^{\dagger} &= \hat{c}_{0\uparrow}^{\dagger}\hat{c}_{0\downarrow}^{\dagger},\\
\hat{P}_{C_1}^{\dagger} &= \hat{c}_{0\uparrow}^{\dagger}\hat{c}_{1\downarrow}^{\dagger},\\
\hat{P}_{C_2}^{\dagger} &= \hat{c}_{0\downarrow}^{\dagger}\hat{c}_{1\uparrow}^{\dagger},\\
\hat{P}_{D_2}^{\dagger} &= \hat{c}_{1\uparrow}^{\dagger}\hat{c}_{1\downarrow}^{\dagger},
\end{aligned}
\end{equation}
the correlated transformation may be written as

\begin{equation}
\label{eq:twoBodyPairRotation}
\hat{P}_{\mu}^{\dagger}
=
\sum_{\nu}
R_{\mu\nu}^{(2)}
\hat{P}_{\nu}^{\dagger},
\end{equation}
where
\[
\nu\in\{D_1,C_1,C_2,D_2\},
\qquad
\mu\in\{D_+,S_0,T_0,D_-\}.
\]

For the two-site system,
\begin{equation}
\label{eq:twoBodyStateRotationMatrix}
\begin{pmatrix}
\ket{D_+}\\
\ket{S_0}\\
\ket{T_0}\\
\ket{D_-}
\end{pmatrix}
=
\frac{1}{\sqrt{2}}
\begin{pmatrix}
1&0&0&1\\
0&1&1&0\\
0&1&-1&0\\
1&0&0&-1
\end{pmatrix}
\begin{pmatrix}
\ket{D_1}\\
\ket{C_1}\\
\ket{C_2}\\
\ket{D_2}
\end{pmatrix}.
\end{equation}
This transformation should be contrasted with the one-body orbital rotation

\begin{equation}
\hat{a}^{\dagger}_{p\sigma}
=
\sum_i R_{ip}\hat{c}^{\dagger}_{i\sigma},
\end{equation}
which transforms individual fermionic modes and automatically induces the corresponding transformation throughout the full Fock space. The correlated transformation above cannot, in general, be represented by such a one-body orbital rotation. Implementing an analogous transformation on a quantum processor would therefore require correlated many-body operations rather than only the fermionic linear-optics transformations used for the orbital rotations considered in this work.

The compact two-dimensional structure of the isolated dimer does not extend directly to the full Fermi-Hubbard ladder. Applying the four-state correlated basis independently to each rung would restrict every rung to its local two-electron sector. Inter-rung hopping, however, transfers particles between neighboring rungs according to processes such as
\[
(N_r,N_{r+1})
\rightarrow
(N_r+1,N_{r+1}-1),
\]
and therefore necessarily accesses local particle-number sectors outside this restricted basis. A rung-local construction containing only
\(\{\ket{D_+},\ket{S_0},\ket{T_0},\ket{D_-}\}\)
would consequently project out physical leg-hopping processes.

An exact rung-local correlated representation could instead retain the complete local Fock space. Because a two-site rung contains four spin orbitals, its full local Hilbert space has dimension \(2^4=16\) and includes the zero-, one-, two-, three-, and four-particle sectors. Such a representation would preserve inter-rung particle transfer, but at the cost of a larger local basis and more complicated transition operators between correlated states. The two-site example therefore demonstrates both the potential advantage and the principal limitation of correlated many-body transformations: they can substantially compress a locally interacting problem, but this compression does not generally extend to the full lattice through the simple, scalable one-body rotations considered in the present work. 

\FloatBarrier

\section{Exact Diagonalization and Compressibility Analysis}

\subsection{Exact Diagonalization Procedure}
\label{sec:EDProcedure}

To avoid the memory cost associated with dense diagonalization, we perform sparse exact diagonalization using an iterative Lanczos eigensolver implemented through the \texttt{scipy.sparse.linalg} submodule~\cite{lanczos1950iteration,lehoucq1998arpack,virtanen2020scipy}. Calculations are carried out within the half-filled \(S_z=0\) symmetry sector for Fermi-Hubbard ladders containing up to \(7\times2\) sites. For each selected orbital basis, the lowest-energy eigenvector is obtained and used to determine the overlap of each occupation-number configuration with the exact ground state. The configurations are then ordered in descending ground-state weight and used to determine the retained determinant subspaces employed throughout the compressibility and exact-reference analyses. This procedure provides both the basis-compressibility results shown in Fig.~\ref{fig:12SiteCompressibility} and the exact-weight-selected reference curves in Fig.~\ref{fig:StatevectorParameterPerformance}.

For each orbital representation considered in the exact-diagonalization analysis, the Fermi-Hubbard Hamiltonian is constructed within the fixed particle-number sector \(N_{\uparrow}=N_{\downarrow}=N/2\). The Lanczos procedure then constructs the Krylov subspace \(\mathcal{K}_{m}\) spanned by

\begin{equation}
\mathcal{K}_{m}(\hat{H},|\mathbf{v}_0\rangle)
=
\operatorname{span}
\left\{
|\mathbf{v}_0\rangle,
\hat{H}|\mathbf{v}_0\rangle,
\ldots,
\hat{H}^{m-1}|\mathbf{v}_0\rangle
\right\},
\end{equation}
and iteratively converges the extremal eigenvalues of the sparse Hamiltonian from an initial trial vector \(|\mathbf{v}_0\rangle\). Only the lowest-energy eigenpair is required for the present analysis, and the resulting ground-state eigenvector \(|\Psi_0\rangle\) is numerically converged using an eigensolver tolerance of \(10^{-10}\).

Once \(|\Psi_0\rangle\) is obtained, its coefficients in the occupation-number basis of the selected orbital representation \(\mathcal{B}\) are used to quantify the relative importance of the corresponding determinants. For a basis configuration \(|\mathbf{n}\rangle_{\mathcal{B}}\), the overlap amplitude is given by

\begin{equation}
\alpha_{\mathbf{n}}^{(\mathcal{B})}
=
{}_{\mathcal{B}}\langle \mathbf{n}|\Psi_0\rangle,
\end{equation}
with the corresponding ground-state weight
\begin{equation}
w_{\mathbf{n}}^{(\mathcal{B})}
=
\left|\alpha_{\mathbf{n}}^{(\mathcal{B})}\right|^2.
\end{equation}
Because the ground-state eigenvector is represented in the same determinant ordering used to construct the Hamiltonian, these weights are obtained directly from the squared magnitudes of its components. The determinants are then sorted in descending order of ground-state weight,

\begin{equation}
w_{(1)}^{(\mathcal{B})}
\geq
w_{(2)}^{(\mathcal{B})}
\geq
\cdots,
\end{equation}
so that the most significant configurations are retained first. For the compressibility calculations, the cumulative sum of these ordered weights is evaluated until a selected threshold \(W\) is reached, and the corresponding number of retained determinants defines \(k_W^{(\mathcal{B})}\).

\subsection{Fixed-Sector Hilbert-Space Dimension}
\label{sec:HilbertSpaceDimension}

At half filling with \(S_z=0\), the fixed particle-number and fixed-spin-sector Hilbert-space dimension for \(N\) lattice sites is \[\dim \mathcal{H}_{N_\uparrow=N_\downarrow=N/2}=\binom{N}{N/2}^2=\left[\frac{N!}{(N/2)!(N/2)!}\right]^2.\] This corresponds to choosing \(N/2\) spin-up electrons and \(N/2\) spin-down electrons among \(N\) spatial sites. For the six-site ladder matrices shown in Fig.~\ref{fig:FHMatrixHamiltonian}, this gives \(\binom{6}{3}^2=400\), resulting in the full \(400\times400\) Hamiltonian matrices, of which the first \(75\times 75\) entries are shown for visual clarity. The size of this fixed-symmetry Hilbert-space can also be approximated by the Stirling approximation: \(\dim \mathcal{H}_{N_\uparrow=N_\downarrow=N/2}\approx(\frac{2}{\pi N})4^{N}\). This asymptotic scaling is used in the system-size analysis associated with Fig.~\ref{fig:12SiteCompressibility}(c,d) in the main text.

\FloatBarrier
\section{DMRG Reference \newline Calculations}
\label{sec:DMRGReference}

Density matrix renormalization group calculations were used to obtain reference ground-state energies for ladder sizes beyond those accessible to exact diagonalization \cite{schollwock2011density, white1992density}. In DMRG, the many-body wave function is represented as a matrix-product state whose accuracy is controlled by the bond dimension \(\chi\). Across a given matrix-product-state bond, the wave function may be expressed through a Schmidt decomposition with singular values \(s_i\). When the number of retained Schmidt states is restricted to \(\chi\), the smaller singular values are discarded, producing the truncation error
\begin{equation}
\epsilon_{\rm trunc}
=
\sum_{i>\chi}s_i^2.
\end{equation}
For a normalized state, \(\epsilon_{\rm trunc}\) therefore represents the fraction of Schmidt weight discarded during the corresponding bond truncation. It should not be interpreted directly as an error in the calculated ground-state energy, but instead provides a diagnostic of how accurately the matrix-product-state representation retains the entanglement structure of the optimized state.

\begin{figure*}[!t]
    \centering
    \includegraphics[width=0.90\textwidth]{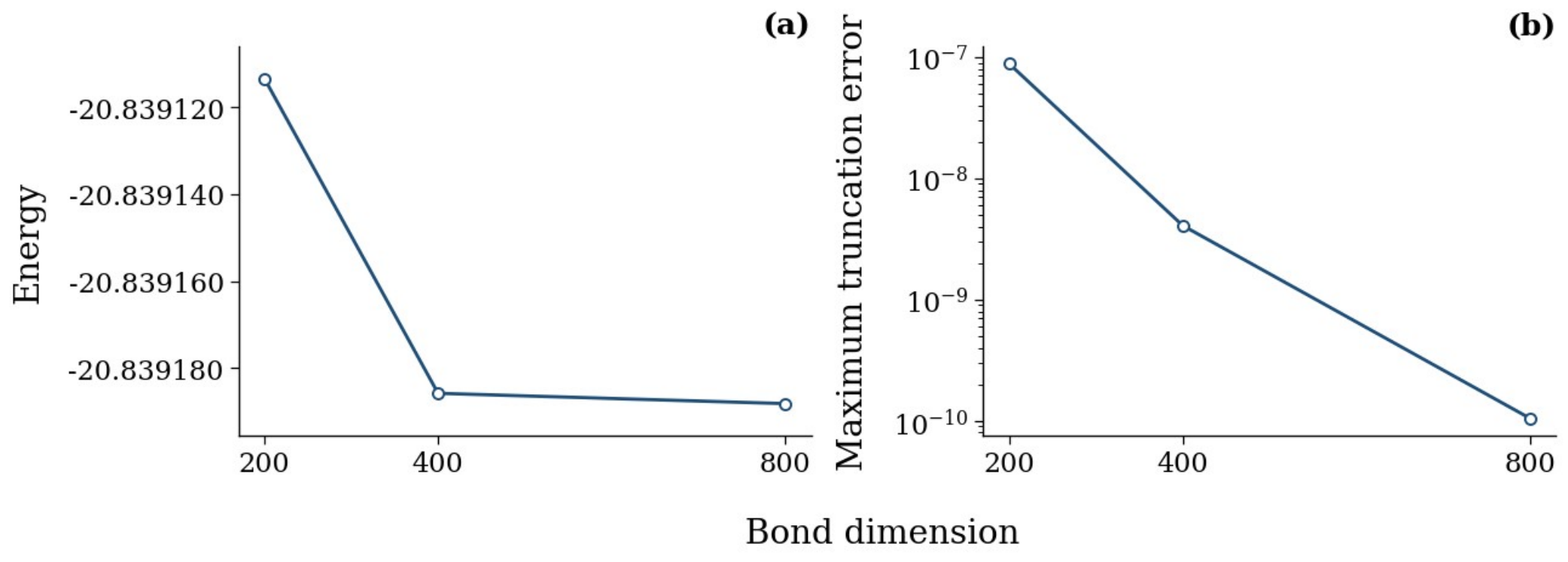}
    \caption{DMRG convergence used to establish the reference ground-state energies. (a) Computed ground-state energy and (b) maximum truncation error as functions of the MPS bond dimension $\chi$. The truncation error corresponds to the discarded Schmidt weight during MPS compression, with the maximum value reporting the largest discarded weight encountered during the DMRG optimization.}
    \label{fig:DMRGTruncationErr}
\end{figure*}

To verify convergence of the DMRG reference calculations, the calculations were repeated while increasing the bond dimension \(\chi\). Figure~\ref{fig:DMRGTruncationErr}(a) shows the resulting ground-state energy as a function of \(\chi\), while Fig.~\ref{fig:DMRGTruncationErr}(b) shows the maximum truncation error encountered during the corresponding DMRG optimization. Reporting the maximum truncation error provides a conservative measure of the largest discarded Schmidt weight occurring during the calculation. As \(\chi\) increases, a larger portion of the Schmidt spectrum can be retained, causing the truncation error to decrease while the calculated energy approaches a stable value. The simultaneous convergence of the ground-state energy and reduction of the discarded weight provides a numerical check on the DMRG reference energies used for comparison with the SKQD calculations in the main text.

\FloatBarrier
\section{Additional Basis-Dependent Compressibility}
\label{sec:AdditionalCompressibility}

\begin{figure*}[!t]
    \centering
    \includegraphics[width=0.88\textwidth]{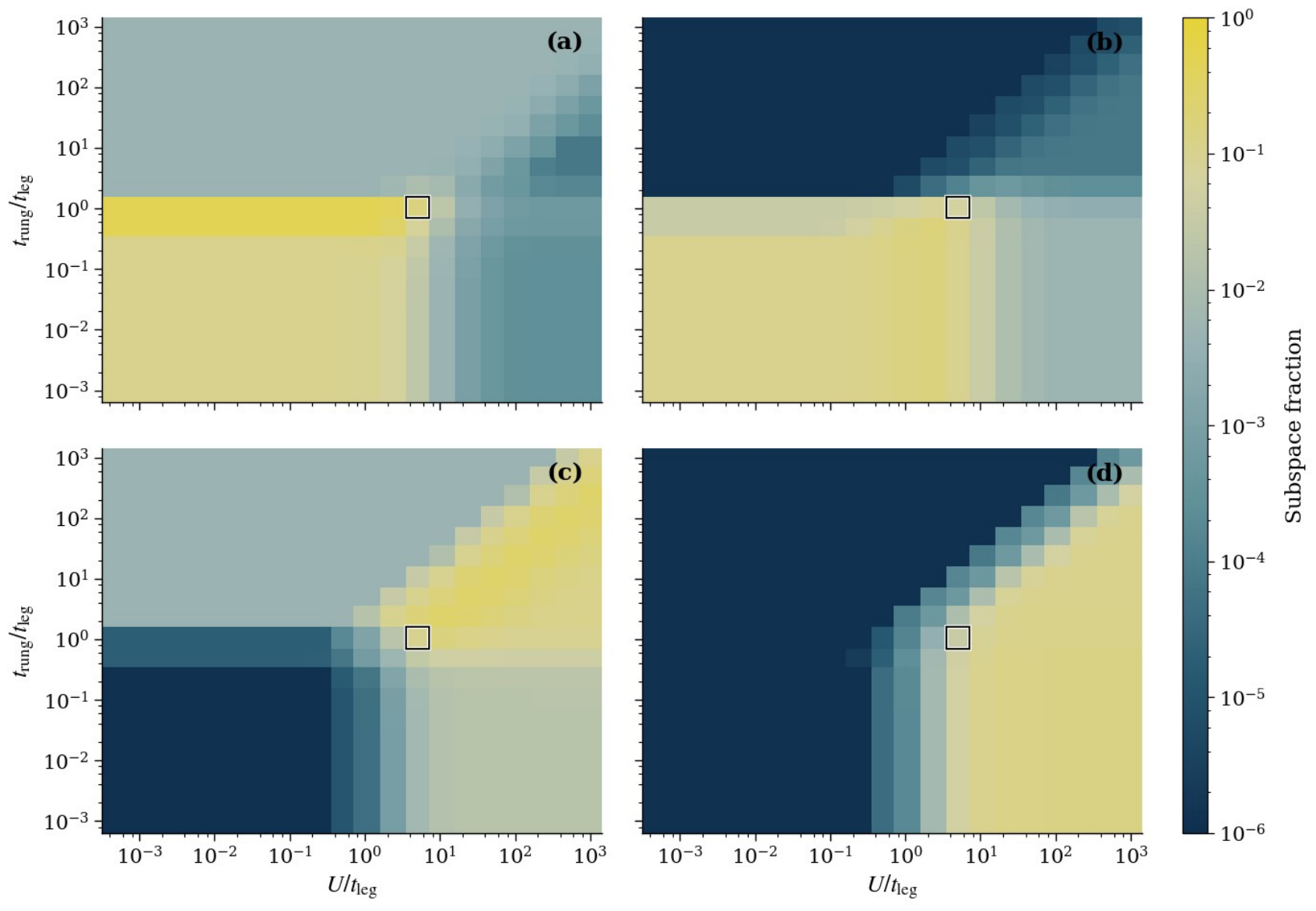}
    \caption{Compressibility of the position (a), bonding/antibonding (b), leg-mode (c), and momentum (d) representations across the sampled parameter space of the 12-site ladder. Compressibility is quantified by the minimum subspace required to retain a cumulative ground-state weight $W\geq0.99$, reported as a fraction of the full half-filled sector. The open squares mark $U/t_{\mathrm{leg}}=5$ and $t_{\mathrm{rung}}/t_{\mathrm{leg}}=1$.}
    \label{fig:CompressibilitybyBasis12Site}
\end{figure*}

Figure~\ref{fig:CompressibilitybyBasis12Site} shows the ground-state compressibility of each of the four orbital representations considered in the main text across the full sampled parameter space of the 12-site Fermi-Hubbard ladder. Whereas Fig.~\ref{fig:12SiteCompressibility}(a) reports only the most compressible representation at each parameter point, the basis-resolved maps shown here illustrate how the required subspace fraction evolves continuously within each representation. The position basis becomes increasingly compact with increasing onsite interaction, the bonding/antibonding representation is favored in the strong-rung regime, and the leg-mode representation becomes more compact when leg hopping dominates. The momentum representation provides the strongest compression in the weakly interacting regime but becomes progressively less compact as correlations increase. The open squares identify the representative intermediate-coupling point used for the system-size scaling analysis in the main text.

\section{SKQD Circuit Construction and Sampling Workflow}
\label{sec:SKQDCircuitPreparation}

Sample-based Krylov quantum diagonalization (SKQD) extends sample-based quantum diagonalization (SQD) by sampling a family of time-evolved states rather than a single reference or variational state. For a Krylov dimension \(K+1\), the sampled states are written as \(\{|\psi_m\rangle\}_{m=0}^{K}\), where \( |\psi_m\rangle=\widetilde{U}(\Delta t)^m|\psi_{\rm ref}\rangle \) and \(\widetilde{U}(\Delta t)\) is a Trotterized approximation to \(e^{-i\hat{H}\Delta t}\). Sampling several Krylov states is intended to enrich the measured configuration space with determinants relevant to the low-energy structure of the Hamiltonian. The effectiveness of this procedure therefore depends both on the overlap of the reference state with the ground state and on the ability of the time-evolved states to preferentially sample configurations carrying significant ground-state weight. If the evolution instead distributes probability broadly across the symmetry-allowed Hilbert space, the sampled subspace can grow substantially without producing a correspondingly efficient ground-state reconstruction.

\begin{figure*}[!t]
    \centering
    \includegraphics[width=0.85\textwidth]{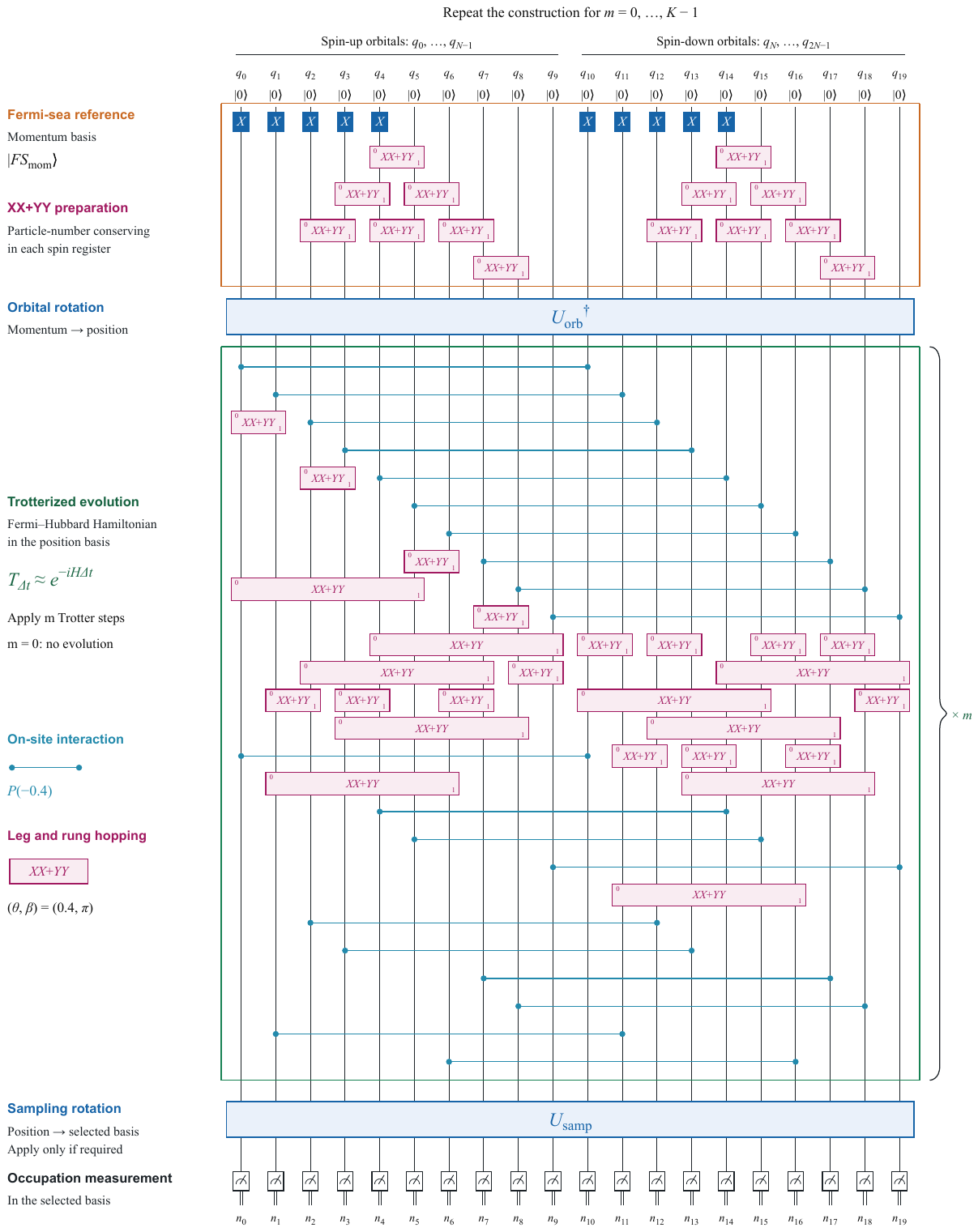}
    \caption{Quantum circuit used for SKQD sampling of the Fermi-Hubbard ladder. A momentum-space Fermi-sea reference state is broadened using particle-number-conserving state-preparation layers and transformed to the position basis using an orbital rotation. Trotterized Fermi-Hubbard evolution is then applied in the position basis, followed, when required, by a final orbital rotation into the selected sampling basis before measurement. Repeating this construction for Krylov indices \(m=0,\ldots,K\) produces the configuration distributions used for reduced-subspace reconstruction.}
    \label{fig:SKQDCircuit}
\end{figure*}

Figure~\ref{fig:SKQDCircuit} summarizes the general circuit construction used in this work; the FSWAP-based implementation of the Trotter-evolution block used for hardware calculations is described separately in Sec.~\ref{sec:FSWAPNetwork}. The reference state is initialized as a momentum-space Fermi sea in which the lowest-energy single-particle modes are occupied according to the desired particle-number and spin sectors. For the half-filled \(S_z=0\) systems considered here, equal numbers of spin-up and spin-down fermions are occupied. Particle-number-conserving \(XX+YY\) state-preparation layers are subsequently applied following the construction of Ref.~\cite{ibm_skqd_tutorial}, broadening the initial Fermi-sea configuration into a controlled superposition while preserving the particle number within each spin sector.

Because the Trotterized Fermi-Hubbard evolution is implemented in the position basis, the prepared momentum-space state is then transformed using the \texttt{ffsim} \texttt{OrbitalRotationJW} operation~\cite{ffsim_orbital_rotation}. This operation implements the corresponding one-body fermionic orbital transformation while preserving particle number. After the orbital rotation, the circuit is expressed in the position basis under the Jordan-Wigner encoding, where the onsite and hopping contributions of the Fermi-Hubbard Hamiltonian can be implemented using local interaction primitives.

The time evolution is constructed using the symmetric second-order Trotter step~\cite{suzuki1990fractal}
\[
\widetilde{U}(\Delta t)
=
e^{-i\hat{H}_U\Delta t/2}
e^{-i\hat{H}_t\Delta t}
e^{-i\hat{H}_U\Delta t/2},
\]
where \(\hat{H}_U\) and \(\hat{H}_t\) denote the onsite-interaction and hopping contributions to Eq.~\eqref{eq:SiteHamiltonian}, respectively. In the position basis, the onsite interaction is diagonal in the occupation-number representation and can be implemented using controlled-phase interactions, while hopping between modes adjacent in the Jordan-Wigner ordering can be implemented using \(XX+YY\)-type interactions. For the hardware calculations, the adapted FSWAP construction described in Sec.~\ref{sec:FSWAPNetwork} is used to bring the required hopping and onsite interactions into local adjacency while reducing explicit Jordan-Wigner parity-string and routing overhead.

For a circuit with Krylov index \(m\), \(m\) Trotter steps are applied following reference-state preparation. The \(m=0\) circuit therefore samples the prepared reference state without Trotter evolution, while the deepest circuit contains \(K\) accumulated Trotter steps and reaches a maximum evolution time \(T_{\max}=K\Delta t\). Unless otherwise specified, the statevector and hardware calculations use
a maximum of \(K=7\) Trotter steps, corresponding to a Krylov dimension \(K+1=8\), with time increments of \(\Delta t=0.025\). Before measurement, an additional \texttt{OrbitalRotationJW} transformation is applied when required to rotate the evolved position-basis state into the selected sampling basis. Position-basis measurements therefore require no additional change of orbital representation after evolution within the quantum circuit, whereas momentum-basis measurements rotate the state back into the momentum representation before measurement.

Each Krylov circuit is sampled independently, and the measured occupation-number configurations are combined to construct the candidate determinant space used in the classical portion of SKQD. Increasing \(K\) or \(\Delta t\) changes the portion of Hilbert space explored by the time-evolved states, while increasing the number of measurement shots improves the sampling resolution of the corresponding probability distributions. As demonstrated by the statevector results and exact-diagonalization reference curves in the main text, broader exploration does not necessarily produce a more efficient reduced subspace if the additional sampled configurations carry comparatively little ground-state weight.

After sampling, the measured bitstrings undergo configuration recovery before construction of the reduced Hamiltonian. The recovery procedure uses the known spin-resolved particle-number constraints of the target Fermi-Hubbard sector to replace or correct samples that fall outside the desired symmetry sector as a result of hardware and measurement errors. The recovered configurations are then used to define a reduced many-body basis, onto which the original Hamiltonian is projected and classically diagonalized. By increasing the retained configuration set, the reconstructed ground-state energy can be examined as a function of subspace dimension. This quantum-sampling and classical-reconstruction procedure constitutes the hybrid SKQD workflow used throughout this work.

\section{FSWAP Network}
\label{sec:FSWAPNetwork}

To minimize routing overhead and parity-string depth on quantum hardware, we adapt the general two-dimensional fermionic SWAP (FSWAP) network demonstrated in Ref.~\cite{alam2025programmable, kivlichan2018quantum} to the ladder geometry studied here. Under the Jordan-Wigner transformation, fermionic modes are mapped to a one-dimensional qubit ordering. Hopping terms between modes that are adjacent in this ordering can be implemented using local \(XX+YY\)-type two-qubit interactions, whereas hopping between nonadjacent modes introduces additional parity strings of \(Z\) operators. These strings encode the fermionic sign associated with exchanging particles, but they also increase circuit depth and routing overhead on hardware with limited connectivity.

For a mode \(n\) in the Jordan-Wigner ordering, the fermionic creation and annihilation operators are mapped as
\begin{equation}
\label{eq:JWDecomposition}
\begin{aligned}
\hat{c}^\dagger_{n}
&=
\left(\prod_{i=1}^{n-1} Z_i\right)
\frac{1}{2}\left(X_n - iY_n\right),\\
\hat{c}_{n}
&=
\left(\prod_{i=1}^{n-1} Z_i\right)
\frac{1}{2}\left(X_n + iY_n\right),
\end{aligned}
\end{equation}
where \(n\) denotes the position of the fermionic mode in the Jordan-Wigner string, and \(X\), \(Y\), and \(Z\) are Pauli operators \cite{tranter2018comparison}. With this mapping, a general hopping term, such as the one described in Eq. \eqref{eq:SiteHamiltonian}, between two modes \(p<q\) takes the form
\begin{equation}
\hat{c}^{\dagger}_{p}\hat{c}_{q}
+
\hat{c}^{\dagger}_{q}\hat{c}_{p}
=
\frac{1}{2}
\left(
X_p Z_{p+1}\cdots Z_{q-1} X_q
+
Y_p Z_{p+1}\cdots Z_{q-1} Y_q
\right).
\end{equation}
Thus, if \(q=p+1\), the intermediate \(Z\)-string is absent and the hopping term reduces to a local \(XX+YY\) interaction. If the modes are separated by \(|q-p|>1\), the hopping term contains \(|q-p|-1\) additional \(Z\) operators.

Generic SWAP gates do not retain the fermionic sign associated with exchanging particles. Instead, FSWAP gates exchange fermionic modes while preserving the antisymmetric sign structure of the encoded fermionic state. In this way, the fermionic sign structure is absorbed into the mode-routing network that already exists within hardware simulation, reducing the need to implement long explicit parity strings for nonlocal hopping terms. The matrix decomposition of an FSWAP operation is as follows:
\begin{equation}
FSWAP = \begin{pmatrix}
    1&0&0&0\\
    0&0&1&0\\
    0&1&0&0\\
    0&0&0&-1
\end{pmatrix}
= CZ\cdot SWAP.
\end{equation}
The FSWAP operation may therefore be decomposed into a SWAP together with a controlled-\(Z\) operation. When compiled to the native gate set of \texttt{ibm\_miami}, the resulting two-qubit interactions are expressed using CZ gates together with the required single-qubit rotations.

\begin{figure*}[!b]
    \centering
    \includegraphics[width=0.75\textwidth]{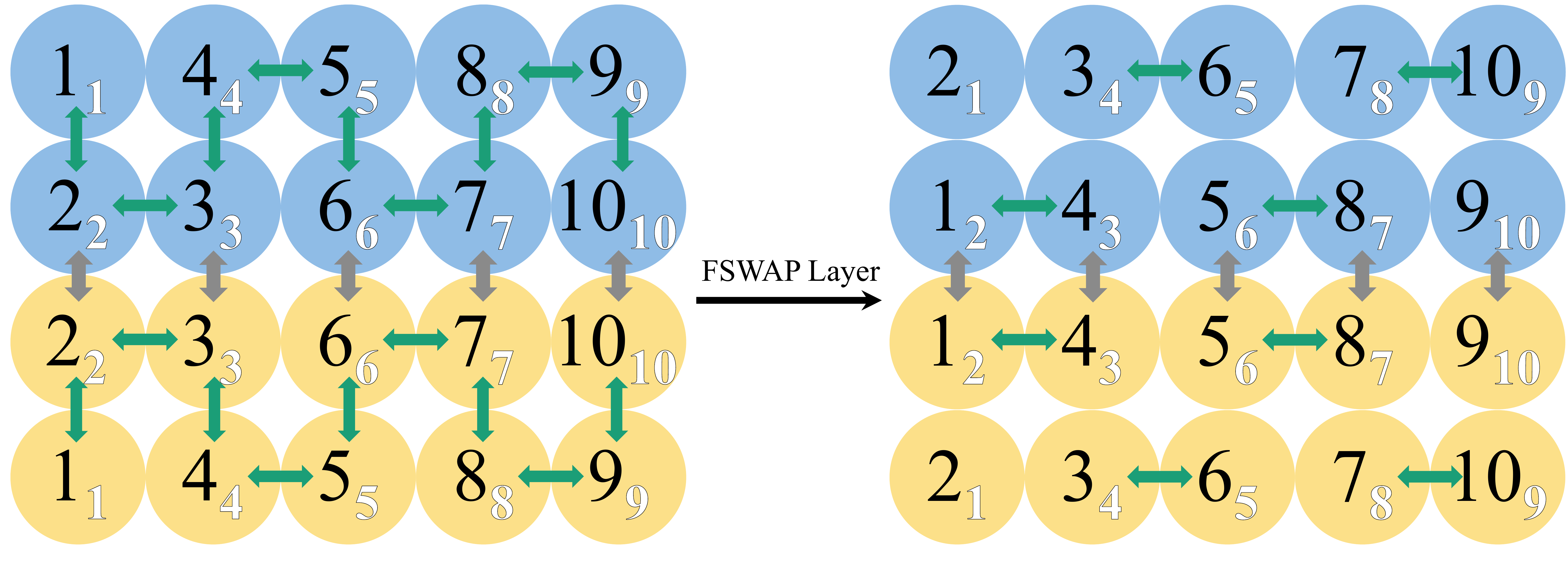}
    \caption{Schematic of a single FSWAP layer used in the folded-vertical circuit construction. Orbital labels show the fermionic ordering before and after the layer, while nearest-neighbor arrows indicate the pairings used to permute the fermionic modes and bring the required interactions into local adjacency.}
    \label{fig:FSWAPLayer}
\end{figure*}

We then integrate an FSWAP network into the Trotterization workflow, following the layout shown in Fig.~\ref{fig:FSWAPLayer}, to model the Fermi-Hubbard ladder on an ideal square-lattice qubit topology where non-edge qubits have four nearest neighbors. The goal of this construction is to reduce the routing overhead and explicit Jordan-Wigner parity-string depth that would otherwise arise from hopping terms between nonadjacent fermionic modes. Under the Jordan-Wigner transformation, fermionic modes are assigned to a one-dimensional ordering, so hopping terms between adjacent modes in the string can be implemented as local \(XX+YY\)-type interactions, while hopping terms between nonadjacent modes require additional \(Z\)-parity operators to preserve the fermionic sign structure. Thus, locality in the physical ladder or corresponding qubit layout does not by itself guarantee locality in the Jordan-Wigner representation; the interacting modes must also be adjacent in the Jordan-Wigner string.

In the initial layout, the Jordan-Wigner ordering is arranged in a zig-zag pattern within each spin sector. For the ladder site labels, this ordering can be viewed schematically as \([1,2,3,4,5,6,\ldots]\), where pairs \([2n-1,2n]\) correspond to vertical rung connections for integer \(n\geq 1\). Adjacent pairs of the form \([2n,2n+1]\) then capture one alternating subset of horizontal leg connections, but the remaining leg connections, such as \([1,4]\), \([3,6]\), and \([5,8]\), are not adjacent in the Jordan-Wigner string even though they are nearby in the qubit layout. Without reordering, these nonadjacent hopping terms would require explicit parity strings. Applying a vertical FSWAP layer across the rung pairs reorders the string to \([2,1,4,3,6,5,\ldots]\), making the complementary set of leg connections adjacent in the Jordan-Wigner ordering. In this way, the FSWAP network allows both subsets of leg hopping terms to be implemented locally at different points within the Trotter step.

A secondary benefit of this mirrored layout is that it also improves the locality of onsite interactions between opposite-spin modes on the same Hubbard site. Before the FSWAP layer, one subset of sites, such as \(2,3,6,7,\ldots\), has spin-up and spin-down modes adjacent across the center of the layout, allowing the onsite interaction to be applied without additional routing. The complementary subset, such as \(1,4,5,8,\ldots\), becomes adjacent after the vertical FSWAP layer. Unlike hopping terms, these onsite interactions do not require Jordan-Wigner parity strings; their challenge is instead hardware connectivity between the two spin sectors. Therefore, by combining the zig-zag Jordan-Wigner ordering with a vertical FSWAP layer, the Trotter step places all required hopping and onsite interactions adjacent either in the corresponding Jordan-Wigner ordering or in the physical qubit layout on an ideal square processor.

\begin{figure*}[!t]
    \centering
    \includegraphics[width=0.9\textwidth]{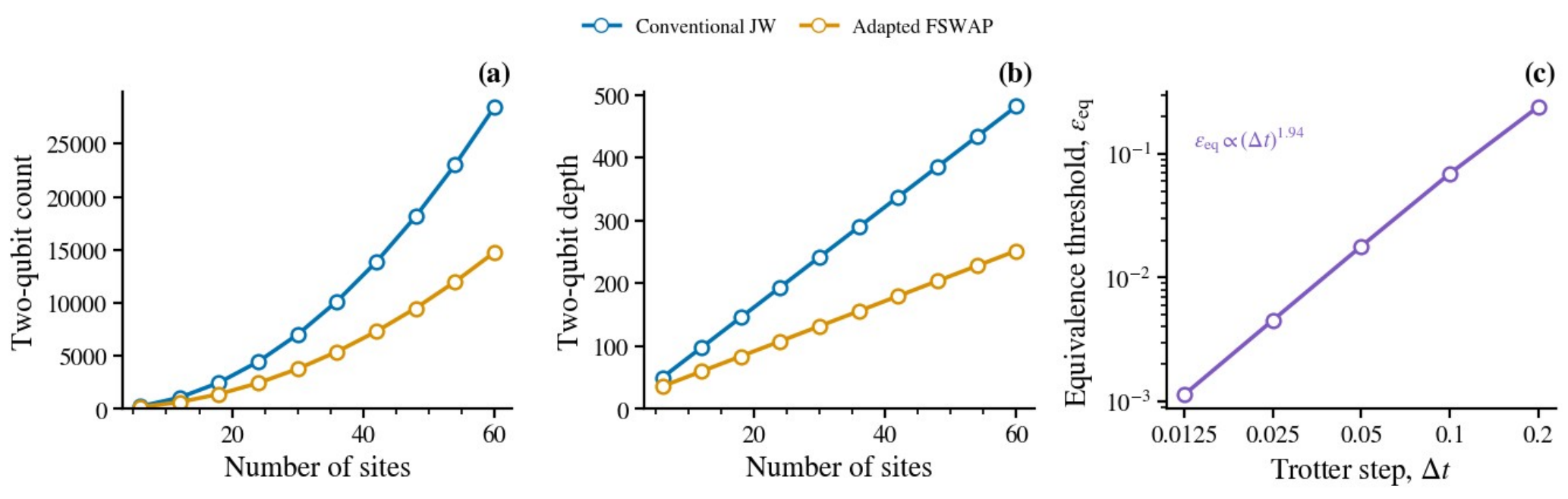}
    \caption{Two-qubit resource scaling and operator-equivalence diagnostics for the adapted FSWAP construction. (a) Two-qubit gate count and (b) two-qubit depth for a single Trotter step using the conventional Jordan-Wigner parity-string construction and the adapted FSWAP construction as functions of the number of lattice sites. (c) Minimum absolute equivalence tolerance, $\varepsilon_{\mathrm{eq}}$, required for the corresponding conventional Jordan-Wigner and adapted-FSWAP circuit unitaries to be identified as equivalent up to a global phase as a function of the Trotter step size $\Delta t$.}
    \label{fig:TranspilationResourceComparison}
\end{figure*}

To investigate the benefit of the adapted FSWAP network, we compare the two-qubit gate count and two-qubit depth of a conventional Jordan-Wigner construction against the FSWAP-based construction described in Fig.~\ref{fig:FSWAPLayer}. Here, two-qubit gate count is defined as the total number of entangling two-qubit operations in the circuit, while two-qubit depth is defined as the minimum number of sequential two-qubit gate layers after gates acting on disjoint qubit pairs are parallelized. In the conventional Jordan-Wigner implementation, hopping terms between nonadjacent modes require additional parity-string operations and, on hardware with limited connectivity, may also require additional routing. Figures~\ref{fig:TranspilationResourceComparison}(a) and \ref{fig:TranspilationResourceComparison}(b) show the scaling of these two constructions with the number of lattice sites in the Fermi-Hubbard ladder. The adapted FSWAP Trotter step exhibits slower growth in both two-qubit gate count and two-qubit depth compared with the conventional Jordan-Wigner parity-string construction.

Figure~\ref{fig:TranspilationResourceComparison}(c) compares the corresponding circuit unitaries for a six-site ladder by plotting the minimum absolute equivalence tolerance, \(\varepsilon_{\rm eq}\), required for the conventional Jordan-Wigner and adapted-FSWAP unitaries to be identified as equivalent up to a global phase. The fitted scaling, \(\varepsilon_{\rm eq}\propto(\Delta t)^{1.94}\), indicates that the discrepancy between the two circuit constructions decreases approximately quadratically with the Trotter step size. This behavior is consistent with the leading difference between the two ordered product formulas entering at \(\mathcal{O}(\Delta t^2)\) \cite{childs2021theory, schubert2023Trotter}.

\begin{figure*}[!b]
    \centering
    \includegraphics[width=0.8\textwidth]{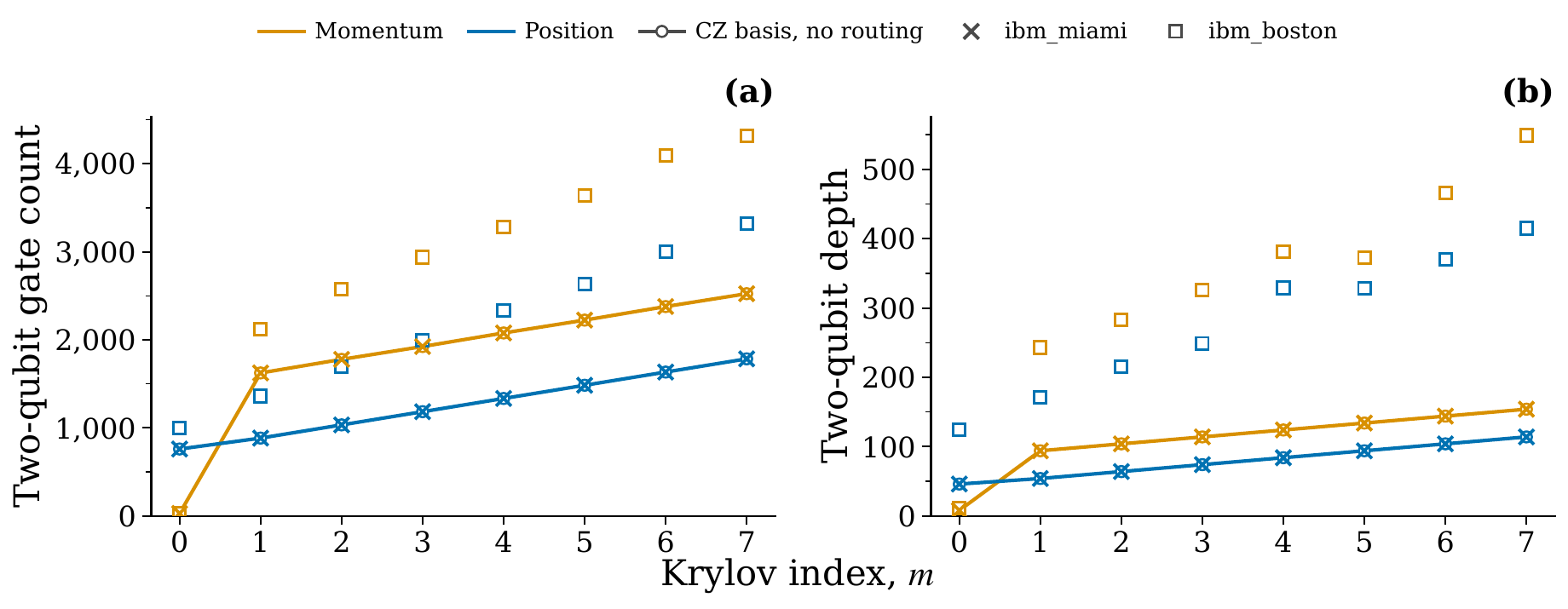}
    \caption{Transpilation resource comparison for the 20-site ladder at $U/t=4$, using three particle-number-conserving state-preparation layers and $\Delta t=0.025$. (a) Two-qubit gate count and (b) two-qubit depth are plotted as functions of Krylov index $m$ for momentum- and position-basis circuits. Crosses and open squares denote transpilation to \texttt{ibm\_miami} and \texttt{ibm\_boston}, respectively, while open circles show the unrouted CZ-basis reference.}
    \label{fig:HardwareResourceComparison}
\end{figure*}

Because the FSWAP permutation is allowed to persist between successive Trotter steps, the correspondence between physical qubits and canonical fermionic-mode labels depends on the Krylov index \(m\). In particular, for position-basis sampling the orbital ordering alternates between the two folded configurations after successive Trotter steps. Consequently, measurement bitstrings from odd-\(m\) position-basis circuits are classically relabeled according to the known FSWAP permutation before being combined with samples from the remaining Krylov circuits. This relabeling restores a common canonical Jordan-Wigner orbital ordering across all sampled circuits without altering the measured probabilities. The consistently labeled configurations can then be supplied to the standard configuration-recovery and reduced-subspace diagonalization procedures used in the SKQD workflow.

\section{Hardware Mapping and Transpilation Resources}
\label{sec:HardwareResources}

\begin{figure*}[!t]
    \centering
    \includegraphics[width=\textwidth]{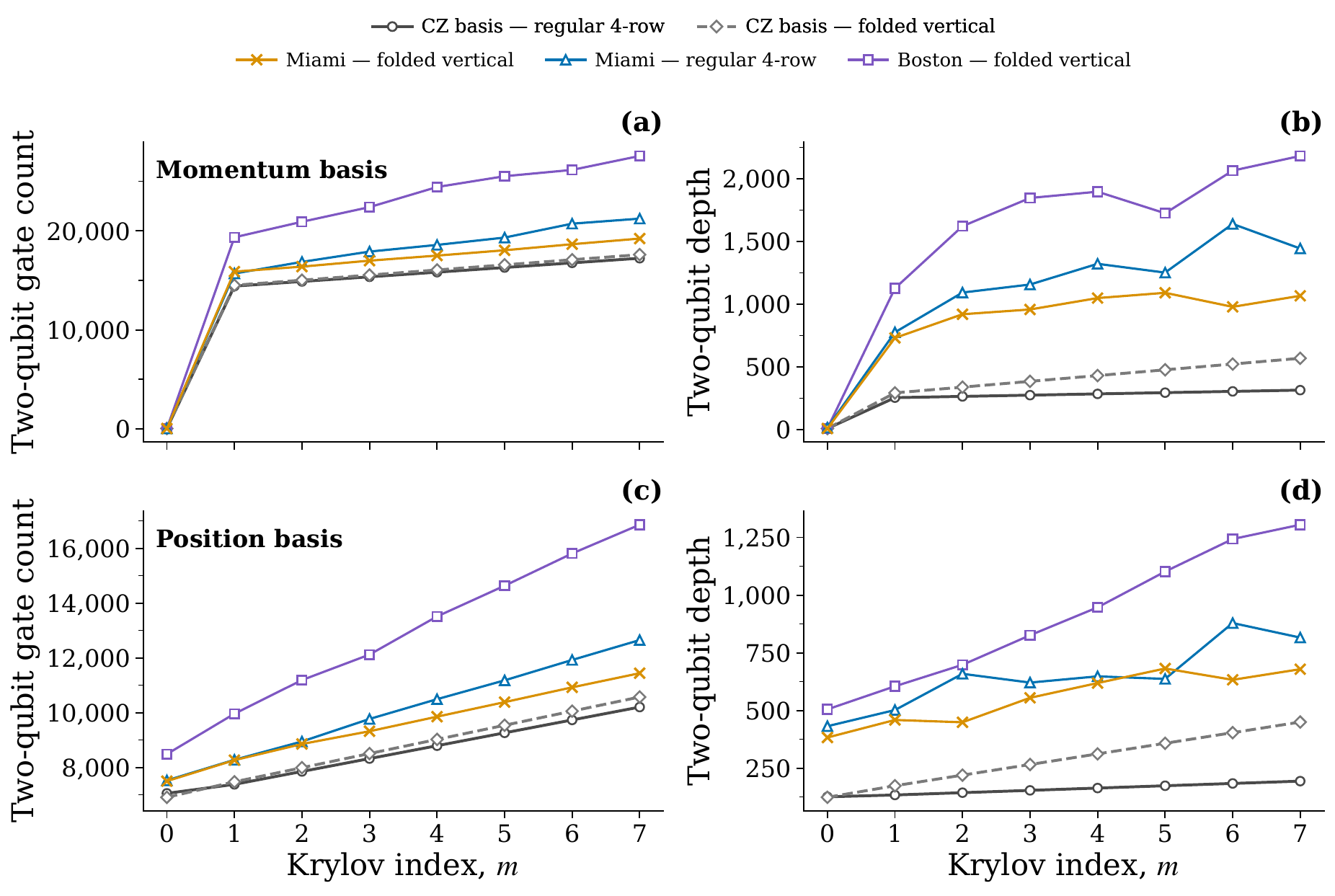}
    \caption{Two-qubit resources for alternative hardware mappings as functions of Krylov index $m$ (this is for 60 sites and 120 qubits). Panels (a,b) show the two-qubit gate count and depth for momentum-basis circuits, while panels (c,d) show the corresponding quantities for position-basis circuits. Curves compare regular four-row and folded-vertical mappings in the unrouted CZ-basis reference and after transpilation on \texttt{ibm\_miami}, with folded-vertical results also shown for \texttt{ibm\_boston}.}
    \label{fig:HardwareResourceVertical}
\end{figure*}

To quantify the hardware overhead associated with the SKQD circuit construction, we compare the two-qubit gate count and two-qubit depth of the resulting circuits under several mapping and transpilation conditions. Figure~\ref{fig:HardwareResourceComparison} shows the resource growth with Krylov index for momentum- and position-basis sampling of the 20-site ladder. As the Krylov index increases, the additional Trotter steps increase both the entangling-gate count and two-qubit depth. Comparison between the unrouted CZ-basis construction and circuits transpiled to \texttt{ibm\_miami} and \texttt{ibm\_boston} separates the intrinsic circuit cost from the additional routing overhead imposed by the physical processor topology.

Figure~\ref{fig:HardwareResourceVertical} further compares the regular four-row and folded-vertical circuit mappings. The folded-vertical construction is designed to place the relevant fermionic modes into local adjacency during the Trotter sequence, thereby reducing the routing overhead introduced when the circuit is transpiled to a fixed-connectivity processor such as \texttt{ibm\_miami}. Although the regular four-row construction requires fewer two-qubit resources in the unrouted CZ-basis reference, this advantage is reduced after the physical connectivity of the processor is imposed. The folded-vertical mapping is better matched to the square hardware topology and produces the lowest transpiled two-qubit resource requirements among the mappings tested for the largest system sizes considered here. These results demonstrate that the logical fermionic ordering and physical qubit mapping jointly determine the practical two-qubit cost of the hardware SKQD circuits.

\FloatBarrier

\raggedbottom
\section{Reconstruction and \newline Sampling Diagnostics}
\subsection{Sensitivity to Omitted Basis States}
\label{sec:OmittedStateSensitivity}

\begin{figure}[!htbp]
    \centering
    \captionsetup{margin=0pt}
    \includegraphics[width=\linewidth]{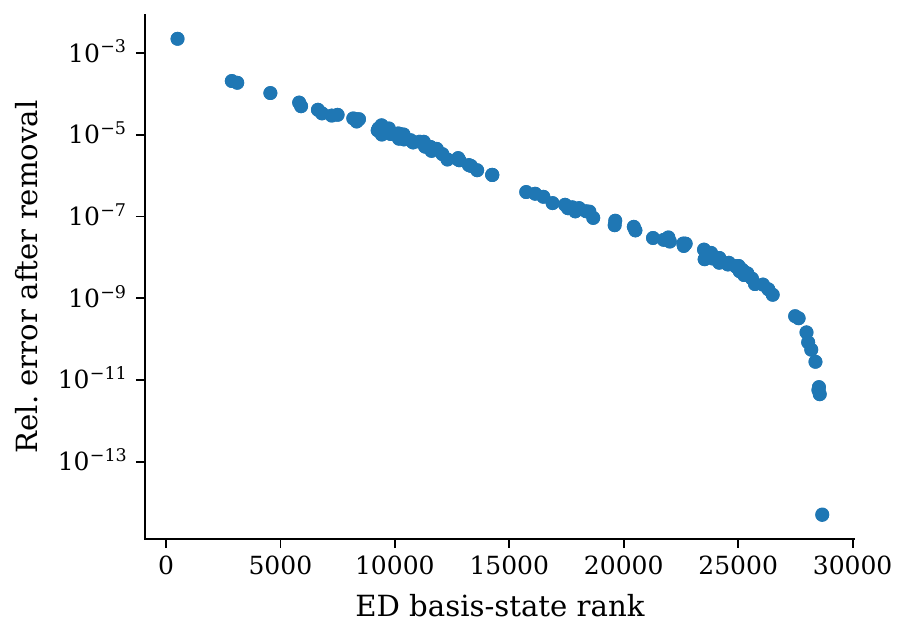}
    \caption{Relative energy error after removal of a single basis state as a function of its rank when configurations are ordered by decreasing exact ground-state probability for the \(U/t=10\) momentum-basis system.}
    \label{fig:SingleStateRemoval}
\end{figure}

The statevector results in Fig.~\ref{fig:StatevectorParameterPerformance}(h) show that a large retained subspace does not necessarily guarantee an accurate reconstructed energy. This behavior is particularly evident for the \(U/t=10\), \(K=7\) momentum-basis calculation, where the reconstructed subspace spans approximately \(99.2\%\) of the fixed particle-number sector while retaining a relative energy error of approximately \(6.3\times10^{-3}\).

To examine the sensitivity of the reconstructed energy to individual configurations, we perform a single-state-removal analysis using the exact-diagonalization reference subspace. Basis states found to be missing from the recovered \(99.2\%\) subspace in Fig.~\ref{fig:StatevectorParameterPerformance}(h) are ordered according to their exact ground-state probability, as determined by exact diagonalization, and each state is removed individually before recalculating the lowest eigenvalue of the projected Hamiltonian. In these calculations, the analyzed subspace then becomes \(D_{full}-1\) as only the single state is removed before carrying out the diagonalization of the new subspace. Figure~\ref{fig:SingleStateRemoval} shows that the sole removal of high-weight basis states produces disproportionately large changes in the reconstructed energy, whereas the removal of low-weight states has progressively smaller influence. The resulting energy sensitivity is strongly correlated with the exact ground-state probability of the removed configuration.

These results demonstrate that reconstruction quality depends not only on the total number of retained configurations but also on the identity of the omitted states. Consequently, even a subspace containing nearly the entire symmetry sector can exhibit a measurable residual error when a small number of configurations important to the low-energy wave function are absent.

\subsection{Sampling-Source Dependence at Fixed Reconstruction Basis}
\label{sec:SamplingDependence}

\begin{figure*}[!t]
    \centering
     \includegraphics[width=0.9\textwidth]{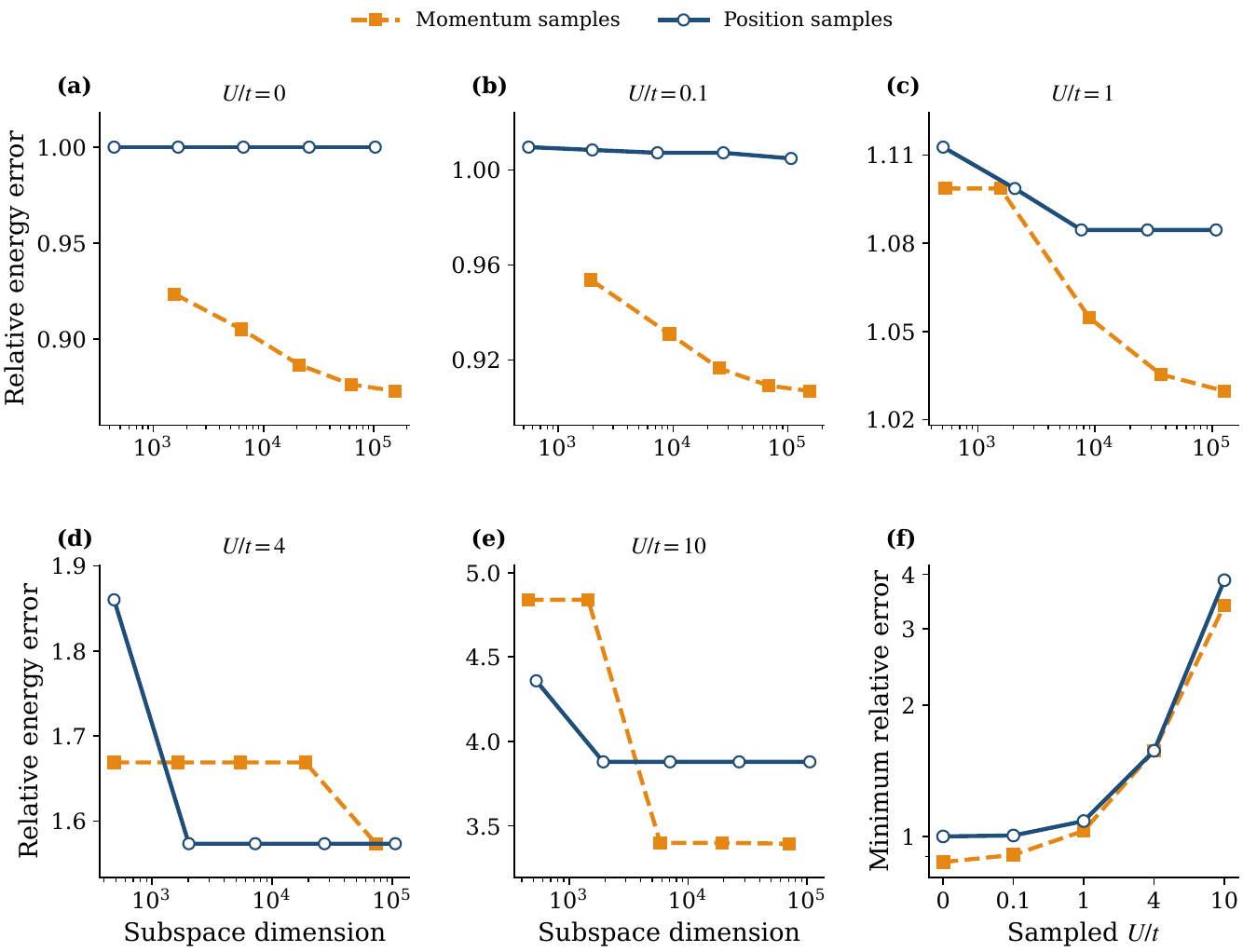}
    \caption{Sampling-source dependence under a common position-basis reconstruction Hamiltonian. The relative energy error is defined as \(\lvert E-E_0\rvert/\lvert E_0\rvert\), where \(E\) is the lowest eigenvalue obtained within the selected reduced subspace and \(E_0\) is the reference ground-state energy obtained from DMRG calculations. Panels (a–e) compare reduced-subspace reconstruction using physical momentum-basis and position-basis hardware samples for the 60-site Fermi-Hubbard ladder at \(U/t=0\), \(0.1\), \(1\), \(4\), and \(10\), respectively. In all cases, the reconstruction Hamiltonian is represented in the position basis, such that only the physical sampling source is varied. Panel (f) summarizes the minimum relative energy error obtained for each sampling source at each interaction strength.}
    \label{fig:SamplingDependence}
\end{figure*}

The hardware results in Fig.~\ref{fig:HardwareEnergy} demonstrate that the quality of a reconstructed SKQD subspace depends not only on its dimension, but also on which configurations are supplied by the quantum sampling procedure. To separate this sampling dependence from the representation used to construct the projected Hamiltonian, we perform an additional controlled comparison in which configurations obtained from physical momentum- and position-basis hardware measurements are both evaluated using the same position-basis Fermi-Hubbard Hamiltonian. In this comparison, the Hamiltonian representation, particle-number constraints, and configuration-recovery procedure are held fixed, while only the physical source of the sampled occupation patterns is changed.

Across the sampled interaction strengths, the configuration sets obtained from momentum-basis measurements produce reconstruction errors that are comparable to or smaller than those obtained from position-basis measurements when both are evaluated with the same position-basis Hamiltonian. The distinction is most apparent in the weakly interacting regime, \(U/t\lesssim1\), where the momentum-sampled configurations consistently yield the lower minimum relative error. Because the projected Hamiltonian is identical for the two sampling sources, this difference cannot be attributed to the intrinsic spectral properties or compressibility of the position-basis Hamiltonian alone. Instead, it demonstrates that the two quantum measurement procedures supply configuration sets of different utility to the subsequent reduced-subspace diagonalization.

This behavior is consistent with the momentum-space structure of the reference-state preparation used in the present SKQD circuit. The circuit is initialized from a Fermi-sea-like configuration in the momentum basis and broadened using particle-number-conserving \(XX+YY\) mixing layers before basis rotations and Krylov evolution. Consequently, the prepared distribution is naturally organized around occupation patterns associated with the kinetic eigenmodes. Transforming this state into the position basis with the \texttt{OrbitalRotationJW} circuit redistributes the amplitudes of these configurations over a larger set of position-basis determinants. In an ideal noiseless calculation this redistribution is a deterministic unitary basis transformation rather than a stochastic process; nevertheless, from the perspective of occupation-number sampling it produces a broader measurement distribution. On quantum hardware, the additional two-qubit gates and accumulated errors associated with this orbital-rotation circuit can further reduce the selectivity of the resulting position-basis samples. This result motivates future investigation of position-adapted reference states, such as a Néel-like state, which may better preserve selective sampling in the position basis.

The comparison in Fig.~\ref{fig:SamplingDependence} therefore provides evidence for sampling-induced configuration spreading that is distinct from basis compressibility itself. In particular, the poorer reconstruction obtained from physical position-basis samples, even when compared against momentum-derived occupation patterns using exactly the same position-basis Hamiltonian, indicates that the position-sampling circuit identifies a less effective subset of determinant indices. The advantage of the momentum-derived samples is consistent with their closer connection to the Fermi-sea-like reference distribution, whereas the position samples are obtained only after the reference amplitudes have been redistributed through the orbital transformation and, for evolved Krylov states, subsequent position-basis dynamics.

\begin{figure}[!htbp]
    \centering
    \captionsetup{margin=0pt}
    \includegraphics[width=\linewidth]{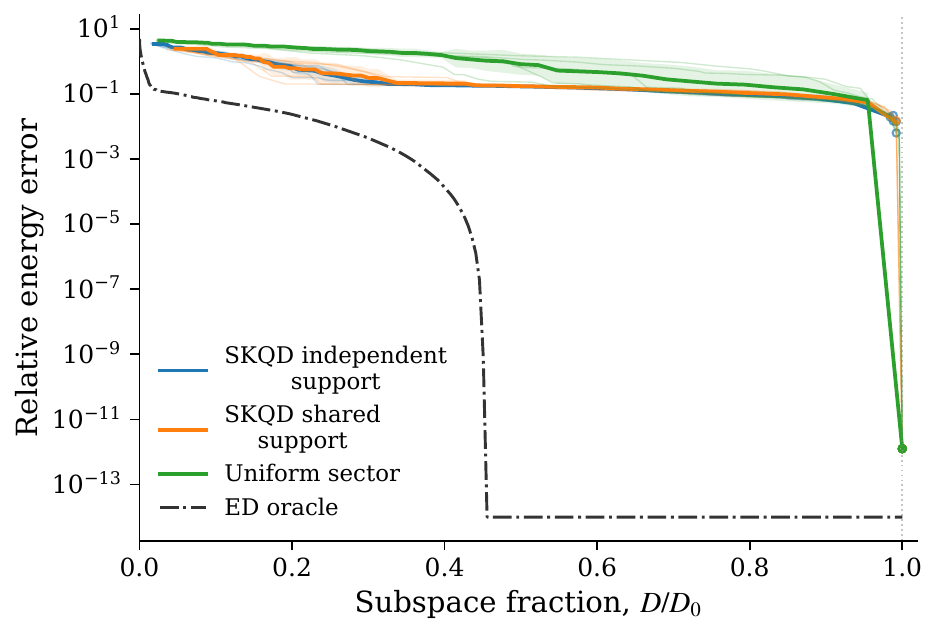}
    \caption{Fulqrum reconstruction for the \(U/t=10\), \(K=7\) momentum-basis statevector dataset using independent spin-sector, shared spin-sector, and uniform-sector sampling. The ED oracle denotes ideal configuration selection by exact ground-state importance.}
    \label{fig:RecoverySamplingComparison}
\end{figure}

This distinction is also important for interpreting the large-system hardware results. A configuration-recovery procedure can enforce the target particle-number sector and increase the number of symmetry-valid determinants without guaranteeing that the recovered set preserves the structure needed for accurate energy reconstruction. The results of Fig.~\ref{fig:SamplingDependence} show that two sampled sets of comparable size can have different reconstruction quality even under an identical Hamiltonian representation. Together with the determinant-connectivity analysis discussed in the main text, this supports the interpretation that successful large-scale SKQD requires both sufficient configuration coverage and a sampling distribution that preferentially retains determinants supporting physically relevant Hamiltonian couplings.

\subsection{Configuration-Recovery \newline Sampling Dependence}
\label{sec:RecoverySamplingDependence}

To further examine the role of configuration construction in the classical recovery stage, we compare Fulqrum reconstructions obtained using independent spin-sector sampling, shared spin-sector sampling, and uniformly sampled configurations from the fixed particle-number sector following the recovery methods utilized in Ref. \cite{robledo2025chemistry, nation2026generalized}. The exact-diagonalization oracle provides a reference in which configurations are retained according to their exact ground-state importance.

Figure~\ref{fig:RecoverySamplingComparison} shows the relative energy error as a function of retained subspace fraction for the \(U/t=10\), \(K=7\) momentum-basis statevector dataset that is also utilized in Figure~\ref{fig:StatevectorParameterPerformance}(h). Both SKQD-derived sampling procedures achieve lower reconstruction errors than uniform sampling over much of the accessible subspace-fraction range, demonstrating that information contained in the quantum sampling distribution remains important after configuration recovery.

Because the spin-up and spin-down sectors are equivalent for the half-filled \(S_z=0\) system considered here, we additionally test a shared spin-sector construction in which the spin-up and spin-down portions of the sampled bitstrings are combined into a common pool before Fulqrum recovery, rather than being treated as two independent sets. No systematic improvement is observed relative to the independent-sector construction over the sampled subspace dimensions. One possible explanation is that the \(K=7\) sampling distribution already provides sufficient spin-resolved support for the Fulqrum reconstruction, such that combining the two pools primarily introduces redundant configurations rather than additional useful support. This interpretation is consistent with the near-saturation behavior observed at large reconstructed dimensions, although a systematic study of the recovery procedure would be required to determine whether this behavior persists across interaction strengths, Krylov depths, and system sizes.

\end{document}